\documentclass{aa}
\usepackage{graphicx}
\usepackage{multirow}

\begin{document}
\title{Energetics of the Kelvin-Helmholtz instability induced by transverse waves in twisted coronal loops}
\titlerunning{Energetics of the KHI induced by transverse waves in twisted coronal loops}
\author{T. A. Howson \and I. De Moortel \and P. Antolin}
\institute{School of Mathematics and Statistics, University of St. Andrews, St. Andrews, Fife, KY16 9SS, U.K.}

\abstract{}
{We quantify the effects of twisted magnetic fields on the development of the magnetic Kelvin-Helmholtz instability (KHI) in transversely oscillating coronal loops.}
{We modelled a fundamental standing kink mode in a straight, density-enhanced magnetic flux tube using the magnetohydrodynamics code, Lare3d. In order to evaluate the impact of an azimuthal component of the magnetic field, various degrees of twist were included within the flux tube's magnetic field.}  
{The process of resonant absorption is only weakly affected by the presence of a twisted magnetic field.  However, the subsequent evolution of the KHI is sensitive to the strength of the azimuthal component of the field. Increased twist values inhibit the deformation of the loop's density profile, which is associated with the growth of the instability. Despite this, much smaller scales in the magnetic field are generated when there is a non-zero azimuthal component present. Hence, the instability is more energetic in cases with (even weakly) twisted fields. Field aligned flows at the loop apex are established in a twisted regime once the instability has formed. Further, in the straight field case, there is no net vertical component of vorticity when integrated across the loop. However, the inclusion of azimuthal magnetic field generates a preferred direction for the vorticity which oscillates during the kink mode.}
{The KHI may have implications for wave heating in the solar atmosphere due to the creation of small length scales and the generation of a turbulent regime. Whilst magnetic twist does suppress the development of the vortices associated with the instability, the formation of the KHI in a twisted regime will be accompanied by greater Ohmic dissipation due to the larger currents that are produced, even if only weak twist is present. The presence of magnetic twist will likely make the instability more difficult to detect in the corona, but will enhance its contribution to heating the solar atmosphere. Further, the development of velocities along the loop may have observational applications for inferring the presence of magnetic twist within coronal structures.}
{}
\maketitle

\section{Introduction}\label{sec:introduction}
With the aid of improved spatial and temporal resolution in contemporary observing instruments, many recent studies have highlighted the abundance of wave energy in the solar corona \citep{Aschwanden1999, Okamoto2007, Tomczyk2007, Parnell2012}. Whether this energy is dissipated sufficiently quickly to significantly contribute to the coronal heating budget remains a contentious issue \citep[e.g.][]{Arregui2015, IDMPC}. However, the potential for waves to drive turbulent flows may have consequences for heating the corona as the small length scales that develop can lead to an increased rate of wave energy dissipation \citep[see e.g. recent work by][]{VanBalle2011, Woolsey2015} .

The rapid damping of transverse kink oscillations in coronal loops has been widely detected in recent years. A number of studies have reported this phenomenon in both standing \citep[e.g.][]{Aschwanden2002, Pascoe2016} and propagating waves \citep[e.g.][]{Verth2010, Morton2014}, and find decay rates far in excess of those expected from typical coronal dissipation. It is widely anticipated that the cause of this enhanced damping is the process of resonant absorption (or mode coupling in the case of propagating waves) through which energy is efficiently transferred from the kink mode to localised Alfv\'enic waves \citep{Ionson1978}. This transfer is caused by the existence of a resonant layer in the loop on which the local Alfv\'en frequency corresponds to the frequency of the transverse wave mode \citep[see][for a comprehensive review]{Goossens2011_Rev}.

This process may be of relevance to the study of coronal heating since the resultant Alfv\'enic waves are subject to dissipation through the process of phase mixing \citep{Pagano2017}. The azimuthal waves are contained within the loop's boundary layer and are associated with large gradients in both the velocity and magnetic fields. The large velocity shear is of particular interest for this paper as it may drive the formation of the Kelvin-Helmholtz instability \citep[e.g.][]{Heyvaerts1983}. Previous analytical studies \citep[e.g.][]{Browning1984, Hollweg1990, Uchimoto1991} and numerical models \citep[e.g.][]{Terradas2008, Antolin2014, Magyar2016, Howson2017} have demonstrated that the KHI may be triggered in oscillating coronal loops. This may have further significance for coronal heating as the formation of the instability generates the possibility of an enhanced rate of wave energy dissipation. In particular, as the KHI develops, a regime of magnetohydrodynamic (MHD) turbulence develops, leading to a cascade of energy to smaller spatial scales and, ultimately, heating.

Whether the KHI is common in the solar atmosphere has not yet been conclusively established. Indeed, the Alfv\'enic waves which may trigger the instability are difficult to detect due to their weak compressibility. Furthermore, they are associated with length scales that reach the spatial resolving power of current observational instruments. Despite this, in the past few years, \citet{Foullon2011, Ofman2011} claim that the appearance of vortices in observations of coronal mass ejections may be evidence of the KHI. In these studies, however, the triggering mechanism is the presence of large plasma velocities parallel to the magnetic field, rather than azimuthal waves. Whilst no direct observations of the KHI forming in transversely oscillating coronal loops have been detected, \citet{Okamoto2015, Antolin2015} interpret plasma behaviour in prominence observations as evidence of resonant absorption and a turbulent regime induced by the KHI. 

Several studies have demonstrated that resonant absorption remains important in various coronal loop configurations. For example, \cite{VanDoors2004, Terradas2006} find that the process is robust in the context of curved loops and \cite{Ruderman2003} considers the process in the case of loops with elliptical cross-sections. Furthermore, studies considering general loop shapes \citep{Pascoe2011} and multi-stranded structuring \citep{Terradas2008b} highlight the insensitivity of resonant absorption (mode coupling) to these conditions. Additionally, studies find that loop stratification due to gravitational effects \citep{Andries2005} and loops containing magnetic twist \citep{Karami2009, Ebrahimi2016} show that these configurations do not prevent the transfer of energy to azimuthal Alfv\'enic modes. Comprehensive reviews of resonant absorption in various geometries are presented in \citet{Ruderman2009} and \citet{Goossens2012}.

Many studies have reported the presence of magnetic field twist in coronal structures that can be observed in prominences \citep[and their subsequent eruption e.g.][]{Rust1996}, in active regions \citep{Demoulin2002} or predicted through numerical simulations of photospheric motions \citep{Torok2003} and flux emergence \citep[e.g.][]{Magara2003, Fan2009}. Indeed, shear flows at the photosphere are expected to introduce some twist into the corona. As a result, investigations into the effects of magnetic twist on the decay of kink mode oscillations are very relevant. It has been suggested that the presence of twist in coronal loops will stabilise the velocity shear through the magnetic tension force and so prevent the formation of the KHI \citep{Soler2010}. On the other hand, weak values of twist may still permit some development of the instability \citep{Terradas2017} and \citet{Zaqarashvili2015} find that standing kink and torsional Alfv\'{e}n waves are always unstable to the KHI close to the antinode at the loop apex. 

We aim to quantify the effect of twist on the non-linear evolution of kink modes by conducting a parameter study on the azimuthal component of the magnetic field. We introduce the numerical model used in Section 2 and describe our results in Section 3. A discussion and some conclusions are then presented in Section 4.

\section{Numerical method}

\subsection{Initial configuration}
\begin{figure}[h]
  \centering
  \includegraphics[width=0.3\textwidth]{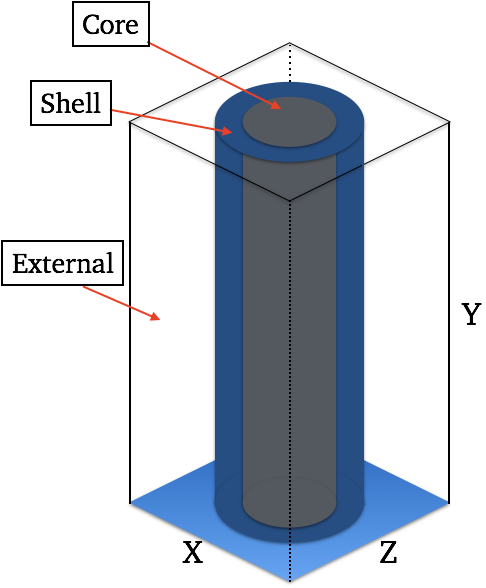}
  \caption{Configuration of the modelled loop.}
  \label{Cartoon_setup}
\end{figure}

\begin{figure}[h]
  \centering
  \includegraphics[width=0.5\textwidth]{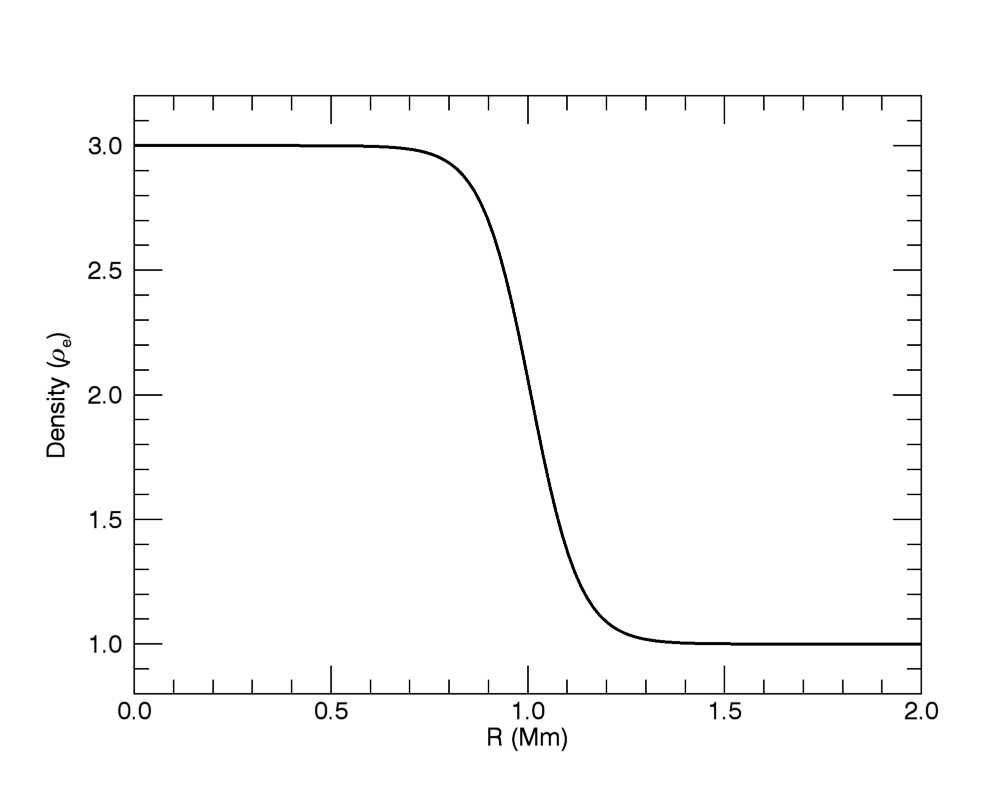}
  \caption{Initial density profile for all cases through the cross-section of the loop. The density is normalised to the initial exterior density, $\rho_e = 8.4 \times 10^{-13}$ kg m$^{-3} $.}
  \label{Initial_Density}
\end{figure}

We followed the work of \citet{Antolin2014, Howson2017} and modelled a coronal loop as a straight flux tube that was denser than the surrounding plasma. The loop was uniform along its length and had a circular cross-section with a dense, inner region (the core) and a boundary layer (the shell) over which the density increased from the external density, $\rho_e = 8.4 \times 10^{-13}$ kg m$^{-3}$ to the internal density $\rho_i=3 \rho_e$. At all heights along the loop, the density was defined as 

\begin{equation} \rho(R) = \rho_e + \frac{(\rho_i-\rho_e)}{2} \left(1 - \tanh\left\{\frac{R - r_a}{r_b}\right\} \right),\end{equation}

where $r_a = 1$ Mm is the loop radius measured to the centre of the boundary layer and $r_b = \frac{1}{16}$ Mm. This ensures the width of the boundary layer is approximately 0.4 Mm. The corresponding density profile across the loop is shown in Figure \ref{Initial_Density}. Measured to the centre of the boundary layer, the radius of the loop was 1 Mm and the length of the loop was 200 Mm. The initial set-up is shown in Figure \ref{Cartoon_setup}. 

Three different cases were considered for the nature of the magnetic field. In the following we describe the set up using cylindrical co-ordinates with the origin at the centre of the loop apex.

\begin{itemize}
\item{\emph{Case 1: Straight field} - A uniform field of strength 21 G aligned with the loop axis ($y$ direction) was included throughout the domain. In this case, in order to ensure the total pressure was uniform, the initial temperature was defined as

\begin{equation} T=\frac{P_0}{\rho}, \end{equation}

where $P_0$ was set to ensure the plasma-$\beta = 0.05$.}
\vspace{0.5cm}

\item{\emph{Case 2: Twisted field; $|B_\phi|$ greatest at core-shell boundary.} - As with the previous case, a magnetic field of strength 21 G was included in the domain. In this case the azimuthal component of the field was defined to be

\begin{equation}
B_{\phi}(R) =  \begin{cases}
           \tau R, \,\,\, &\text{if $R \le a$}, \\
           \tau(b-R), \,\,\,  &\text{if $ a < R \le b$}, \\
          0, \,\,\, &\text{if $R> b$}. \\
\end{cases}
\end{equation}

Here, $\tau$ is a parameter quantifying the level of twist, $a$ is the radius of the core region and $b$ is the radius of the loop (including the boundary region). The profile of $B_{\phi}$ is constant with height and we note that if $\tau=0$, the first case is recovered. Three cases of twist were considered, $\tau = \left\{ 0.1, 1.0, 2.0 \right\}$ and are labelled as Case 2a, 2b and 2c, respectively. The amount of twist that these values produce is presented in Table 1.

In order to maintain a constant field strength, we stipulated that the vertical magnetic field is 

\begin{equation}
\label{By}
B_y^2 (R) = B_0^2 - B_\phi^2(R),
\end{equation}

where $B_0 = 21$ G. In this geometry, the radially inwards tension force, $T_R$ is given by 

\begin{equation} T_R = \frac{B_{\phi}^2}{R}. \end{equation}

Therefore, in order to balance the force associated with the magnetic twist, the gas pressure was defined as 

\begin{equation}
P = \int \frac{B_\phi ^2}{R} \, \text{d} R,
\label{pressure}
\end{equation}
and the constant of integration was selected to ensure the plasma-$\beta = 0.05$.

}
\vspace{0.5cm}

\item{\emph{Case 3: Twisted field; $|B_\phi|$ greatest at shell-exterior boundary. } - Once again, a magnetic field of strength 21 G was included throughout the domain. In this case the azimuthal component of the field was defined to be 

\begin{equation}
B_{\phi}(R) =  \begin{cases}
           \psi R, \,\,\, &\text{if $R \le b$}, \\
           \psi(2b-R), \,\,\,  &\text{if $ a < R \le 2b$}, \\
          0, \,\,\, &\text{if $R> 2b$}. \\
\end{cases}
\end{equation}

Here, the twist extends beyond the density-enhancement of the loop. This case was included in order to determine whether the location of twist within a loop affected the development of the KHI. Only $\psi=1.0$ was considered and the corresponding level of twist is shown in Table 1.

Once again, in order to maintain an initial equilibrium, the vertical magnetic field was given by equation \ref{By} and the gas pressure was defined as in equation \ref{pressure}.}
\end{itemize}

\begin{figure}[h]
  \centering
  \includegraphics[width=0.5\textwidth]{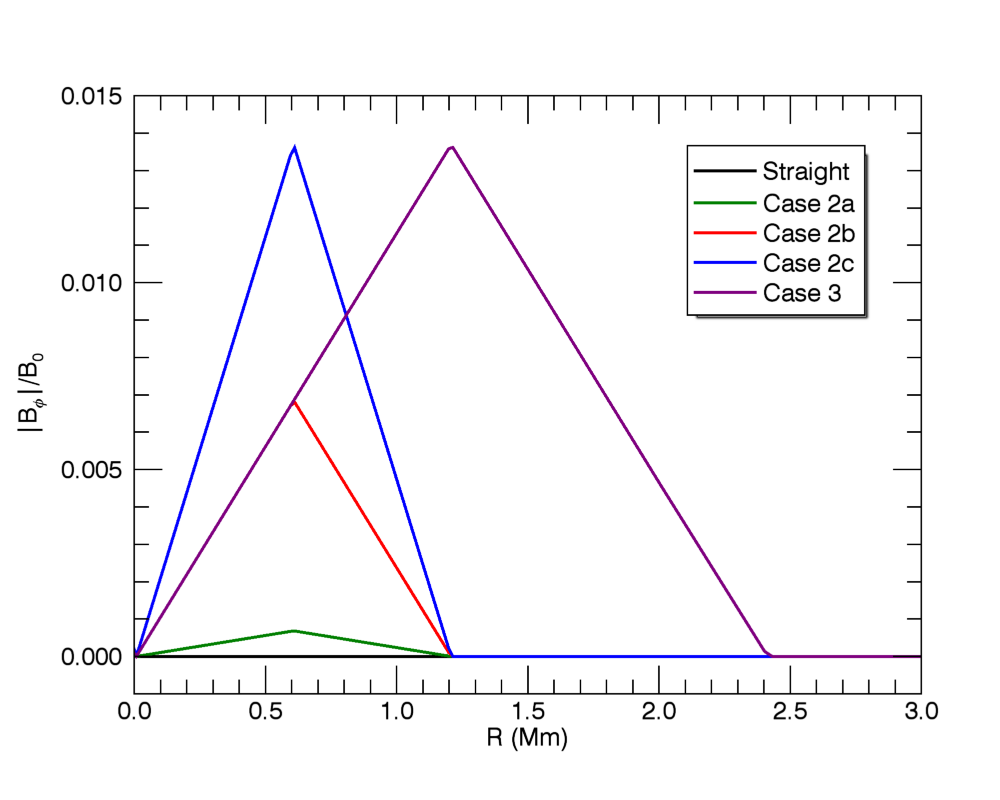}
  \caption{Magnitude of initial azimuthal magnetic field, $|B_\phi|$, through the cross-section of the loop for each case. Here, $B_\phi$ is normalised to $B_0 = 21$ G. We see that in all cases, $|B_\phi| \ll |B_0|$.}
  \label{Bphi}
\end{figure}

\begin{figure}[h]
  \centering
  \includegraphics[width=0.5\textwidth]{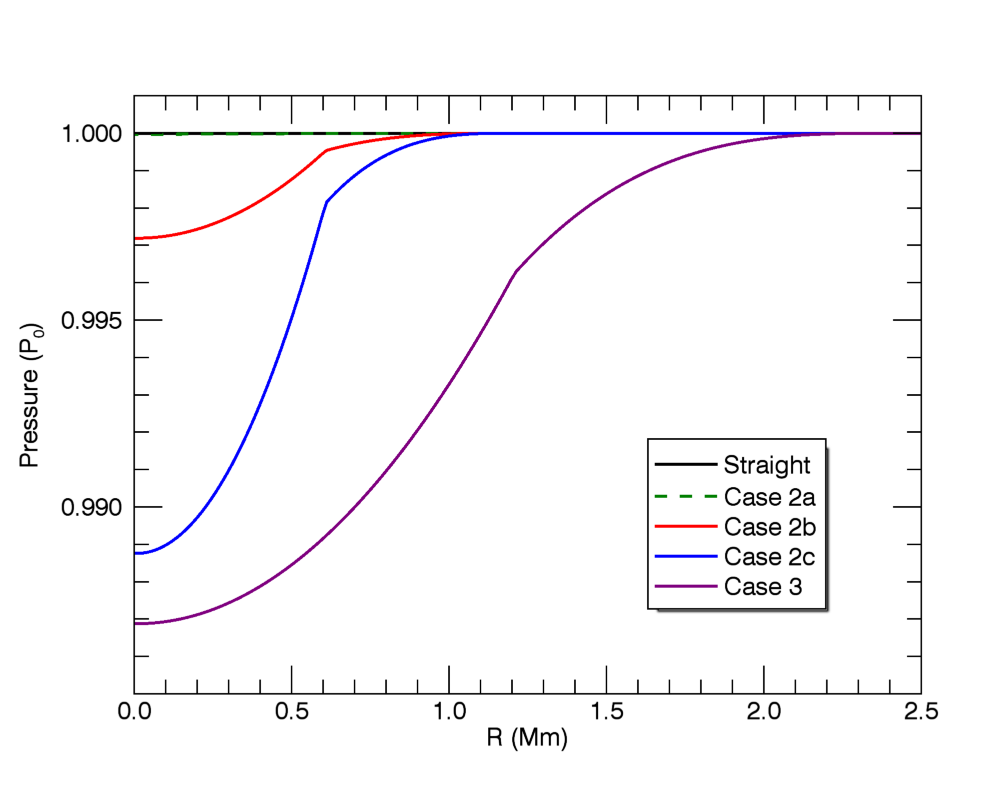}
  \caption{Initial gas pressure profile through the cross-section of the loop for each case. Here, we have normalised the pressure to the external pressure, $P_0$.}
  \label{Pressure}
\end{figure}

Since in each of the three cases, $B_R = 0$ and there is azimuthal and vertical invariance, we immediately see that $\nabla \cdot \vec{B} = 0$ is satisfied. 

We quantified the amount of twist by determining the angle a field line moves through between the lower and upper boundaries of the domain. The cases considered are displayed in Table 1. The initial profiles of the azimuthal field strength and the pressure are shown in Figures \ref{Bphi} and \ref{Pressure}, respectively. We see that even in the most twisted configurations, the gas pressure inside the loop is always above $98\%$ of the external gas pressure. In all cases the temperature profiles are very similar and decrease from an external temperature of 2.5 MK to an internal temperature of around 0.8 MK. Such a profile may be consistent with a loop in thermal non-equilibrium during its cooling phase.

Following \citet{Howson2017} (which we will henceforth refer to as Paper 1), in each case, the fundamental kink mode was excited using the following velocity profile

\begin{equation*} v_x =A \left(\rho - \rho_e \right) \cos \left( \frac{\pi y}{2 L}\right),\end{equation*}
\begin{equation} v_y =0, \end{equation}
\begin{equation*} v_z =0. \end{equation*}

Here, $L=200$ Mm is the length of the modelled loop and the constant, $A$, is selected to produce a maximum velocity at the loop apex $\left(y=0\right)$ of 8.3 kms$^{-1}$. This oscillation corresponds to commonly observed low amplitude kink modes \citep[e.g.][]{Anfino2015}.

\begin{table}[]
\centering
\caption{Twist parameter space.}
\begin{tabular}{c|c|c}
Case & Field      & Angle ($^{\circ}$) \\ \hline
1    & Straight   & 0            \\ \hline
2a   & $\tau=0.1 $  & 0.39         \\
2b   & $\tau = 1.0 $& 39           \\
2c   & $\tau=2.0 $  & 78           \\ \hline
3    & $\psi=1.0$   & 78          
\end{tabular}
\end{table}

\subsection{Numerical method}
In each case, we numerically advanced the full, 3-D, resistive MHD equations in normalised form using the {\bf{La}}grangian-{\bf{re}}map code, Lare3D \citep{Larey}. In order to avoid numerical instabilities, the code uses a staggered grid and implements an adaptive time step restriction so as not to violate the CFL condition whilst maintaining calculation efficiency. The equations are given by

\begin{equation}\frac{\text{D}\rho}{\text{D}t} = -\rho \vec{\nabla} \cdot \vec{v}, \end{equation}
\begin{equation}\label{motion} \rho \frac{{\text{D}\vec{v}}}{{\text{D}t}} = \vec{J} \times \vec{B} - \vec{\nabla} p + \vec{F}_\nu, \end{equation}
\begin{equation}\label{energy}\rho \frac{{\text{D}\epsilon}}{{\text{D}t}} = \eta |\vec{J}|^2 - p(\vec{\nabla} \cdot \vec{v}) + H_\nu, \end{equation}
\begin{equation}\label{induction} \frac{\text{D}\vec{B}}{\text{D}t}=\left(\vec{B} \cdot \vec{\nabla}\right)\vec{v} - \left(\vec{\nabla} \cdot \vec{v} \right) \vec{B} - \vec{\nabla} \times \left(\eta \vec{\nabla} \times \vec{B}\right). \end{equation}

Here, variables have their usual meanings and the non-ideal contributions arise from the inclusion of viscous forces, $\vec{F}_\nu$, in the equation of motion (\ref{motion}), the associated heating term $H_\nu$ in the energy equation (\ref{energy}) and the resistivity $\left( \eta \right)$-dependent terms in (\ref{energy}) and the induction equation (\ref{induction}). The viscous force, $\vec{F}_{\nu}$, and viscous heating term, $H_{\nu}$, are given by 

\begin{equation*}\vec{F}_\nu = \nu \left(\nabla^2 \vec{v} + \frac{1}{3} \nabla \left(\nabla \cdot \vec{v}\right)\right),\end{equation*}
\begin{equation*}H_\nu = \nu \left( \frac{1}{4} \epsilon_{i,j}^2 - \frac{2}{3} \left( \nabla \cdot \vec{v}\right)^2\right),\end{equation*}

where

\begin{equation*} \epsilon_{i,j} = \frac{\partial v_i}{\partial x_j} + \frac{\partial v_j}{\partial x_i}. \end{equation*}

For comparison purposes, the domain used was identical to the high resolution case presented in Paper 1. In all cases, the numerical domain had dimensions of 64 Mm $\times$ 200 Mm $\times$ 64 Mm and used 512 $\times$ 100 $\times$ 512 grid points. The loop was aligned with the $y$-direction and the initial velocity pertubation was aligned with the $x$-direction. The origin $\left(x=y=z=0\right)$ was located at the centre of the loop apex.

In order to ensure a standing mode was generated, the velocities were forced to zero on the $y$ boundaries. All other variables have zero gradients on the edges of the domain. We note that since the magnetic twist is weak ($B_{\phi} \ll B_y$) in all of our simulations, this is sufficient to ensure a standing mode is generated. However, in cases with more significant twist, it may become necessary to use line-tying boundary conditions.

In order to minimise the boundary effects, a non-uniform grid was implemented, with decreasing spatial resolution in the $x$-$z$-plane away from the centre of the domain. To limit any effects of this non-uniformity, the grid was designed so that the oscillation and any subsequent fine scale dynamics occurred within a uniform high resolution region. The cell resolution in the centre of the domain corresponded to 15.9 km $\times$ 2000 km $\times$ 15.9 km. The $y$-axis had a uniform spatial resolution. 

All simulations were essentially ideal since the values of resistivity, $\eta = 10^{-20}$ and viscosity, $\nu = 10^{-20}$, are much smaller than the estimated numerical values. For the spatial resolution used, the numerical (and effective) values of $\eta$ and $\nu$ are of the order $10^{-5}-10^{-6}$ as discussed in Paper 1.

\section{Results}
In all cases, the initial velocity profile induces the fundamental, standing kink mode with a maximum displacement at the loop apex of approximately 0.27 Mm (less than the loop radius of $\approx 1$ Mm) and a period of approximately 280 s. The maximum displacement at each height is proportional to the maximum initial velocity at that height.

\begin{figure}
  \centering
  \includegraphics[width=0.5\textwidth]{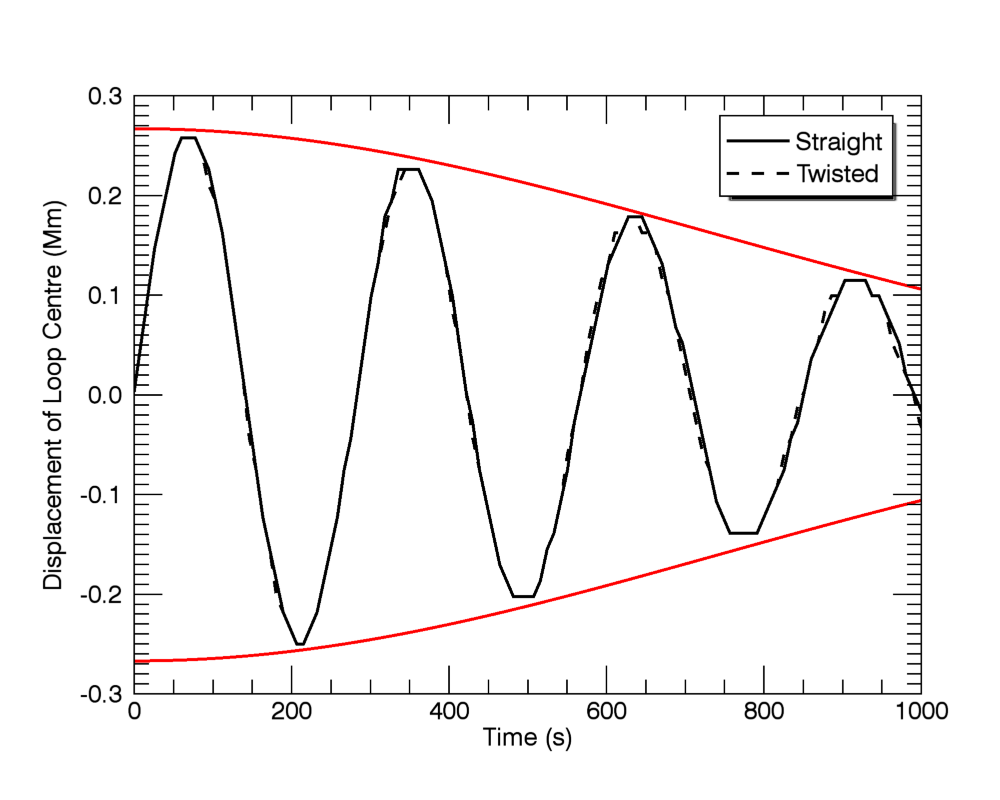}
  \caption{Displacement of the loop apex through time for the straight case and a twisted case (Case 2b). The red lines are given by $y = \pm 0.267 e^{-\left(\frac{t}{1040}\right)^2}$.  }
\label{Loop_Damping}
\end{figure}

\subsection{Resonant absorption}
The displacement of the loop apex as a function of time for both the straight case and a twisted case (2b) is shown in Figure \ref{Loop_Damping}. We observe that the damping rate is largely independent of the twist present and follows a Gaussian profile (red lines). This is in accordance with previous studies that have explained such a (initial) damping rate for both standing \citep{Ruderman2013} and propagating \citep{Gauss_Damp1} waves. This damping is an ideal process and is caused by the transfer of energy from the kink wave to azimuthal, Alfv\'enic modes contained within the loop's boundary region through the well studied process of resonant absorption. 

A resonance exists within the loop's shell region at locations at which the kink frequency coincides with the local Alfv\'en frequency. In the straight case this is the location on which the density, $\rho = \frac{\rho_i + \rho_e}{2} = 2$. However, in the twisted cases, the increased length of the magnetic field lines causes the Alfv\'en frequency to be lower at this location. Consequently, the resonant layer moves to a new location further from the loop centre. However, the existence of any resonant layer is sufficient for the transfer of energy to continue \citep{Terradas2008b, Pascoe2011} and so there is little change to the actual damping profile.

\begin{figure}[h]
  \centering
  \includegraphics[width=0.5\textwidth]{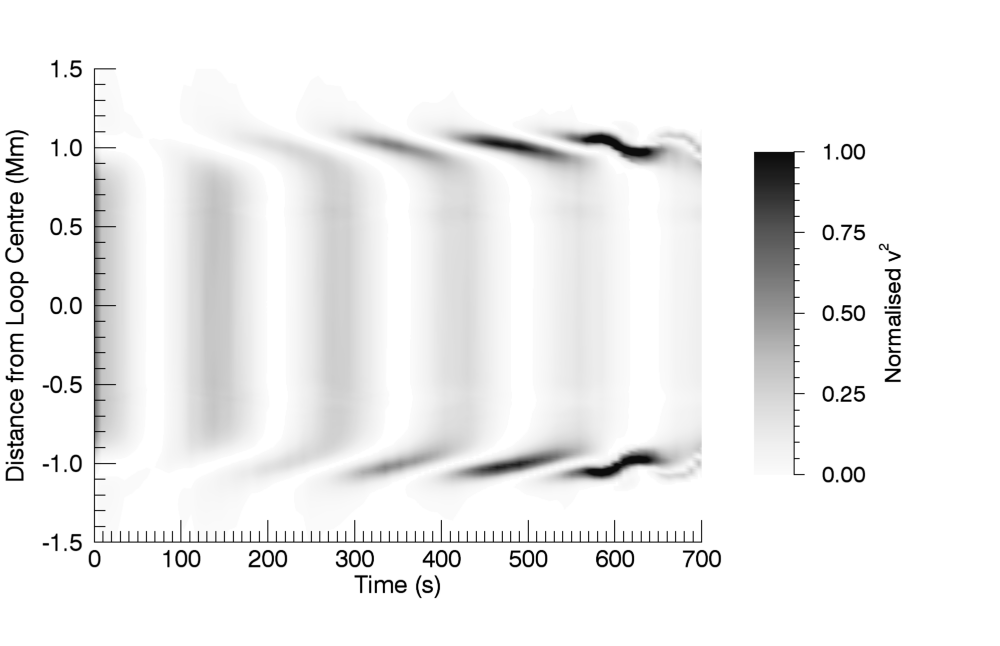} 
  \caption{Evolution of the square of the velocity along the line $y=0$ (through the loop apex), $z=0$. Here $v^2$ is normalised to the maximum value obtained during the simulation. The figure is produced from Case 2b, a twisted field simulation.}
  \label{Res_abs}
\end{figure}

In Figure \ref{Res_abs} we highlight the concentration of flows in the boundary layer as the kink mode decays. We show the evolution of $v^2$ in the line $x=xc, y=0$ where $xc(t)$ is the location of the loop centre. This is a line perpendicular to the initial velocity in the horizontal cross-section at the loop apex. The observed oscillatory pattern arises from the periodic transfer of magnetic to kinetic energy associated with the kink wave. The nature of the initial velocity profile dictates that the majority of the kinetic energy is within the core region of the loop. However, as the kink mode couples with the azimuthal wave modes, energy is transferred into the loop's shell region (approximately 0.8 Mm and 1.2 Mm from the loop centre). As this mode conversion proceeds, the oscillatory pattern survives; however, the length scales of the velocity field are now much smaller than previously. In non-ideal simulations, this would increase the rate of viscous dissipation but can also cause the loop to become Kelvin-Helmholtz unstable (see below). Although Figure \ref{Res_abs} displays the effects of resonant absorption for a twisted loop, a similar figure produced for the straight case is almost identical.

\begin{figure}[h]
  \centering
  \includegraphics[width=0.5\textwidth]{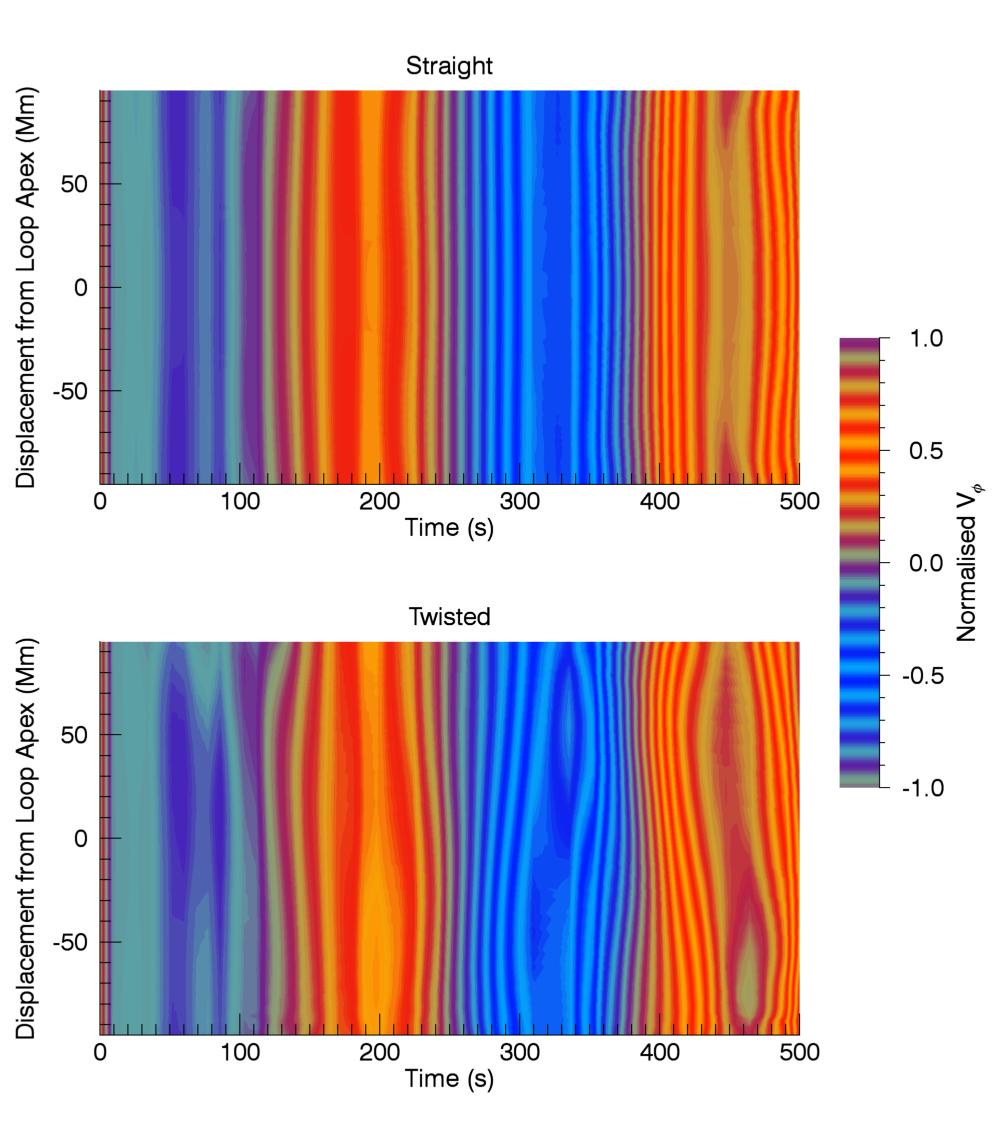}
  \caption{Time evolution of the azimuthal component of the velocity $v_\phi$, along a line parallel to the loop axis within the shell region. The velocity is normalised to the maximum initial velocity at that height. The upper panel shows the simulation with a straight magnetic field (Case 1) and the lower panel shows Case 2c.}
  \label{Harmonics}
\end{figure}

Whilst the quantity of energy transferred from the kink mode to the Alfv\'enic modes is largely independent of the magnetic twist, the presence of an azimuthal component in the field excites additional harmonics of the azimuthal Alfv\'enic waves. In particular, the magnetic twist enhances the non-linearity of the system which is known to produce fluting modes \citep{Ruderman2010, Ruderman2014, Terradas2017}. In Figure \ref{Harmonics}, we show the time variation of $v_\phi$ along the line 

\begin{equation*} x=c_x(y, t), \end{equation*}
\begin{equation*} z=c_z- \zeta, \end{equation*}

 where $c_x(y, t)$ and $c_z$ are the $x$ and $z$ coordinates of the loop centre, respectively and $\zeta = 1$ Mm. This line moves with the global kink mode and is always contained within the loop boundary. It lies on the resonant layer in the straight case where the amplitude of the Alfv\'enic waves is expected to be greatest. For each height, we have normalised $v_\phi$ by the maximum initial velocity at that height.

\begin{figure*}
  \centering
  \includegraphics[width=\textwidth]{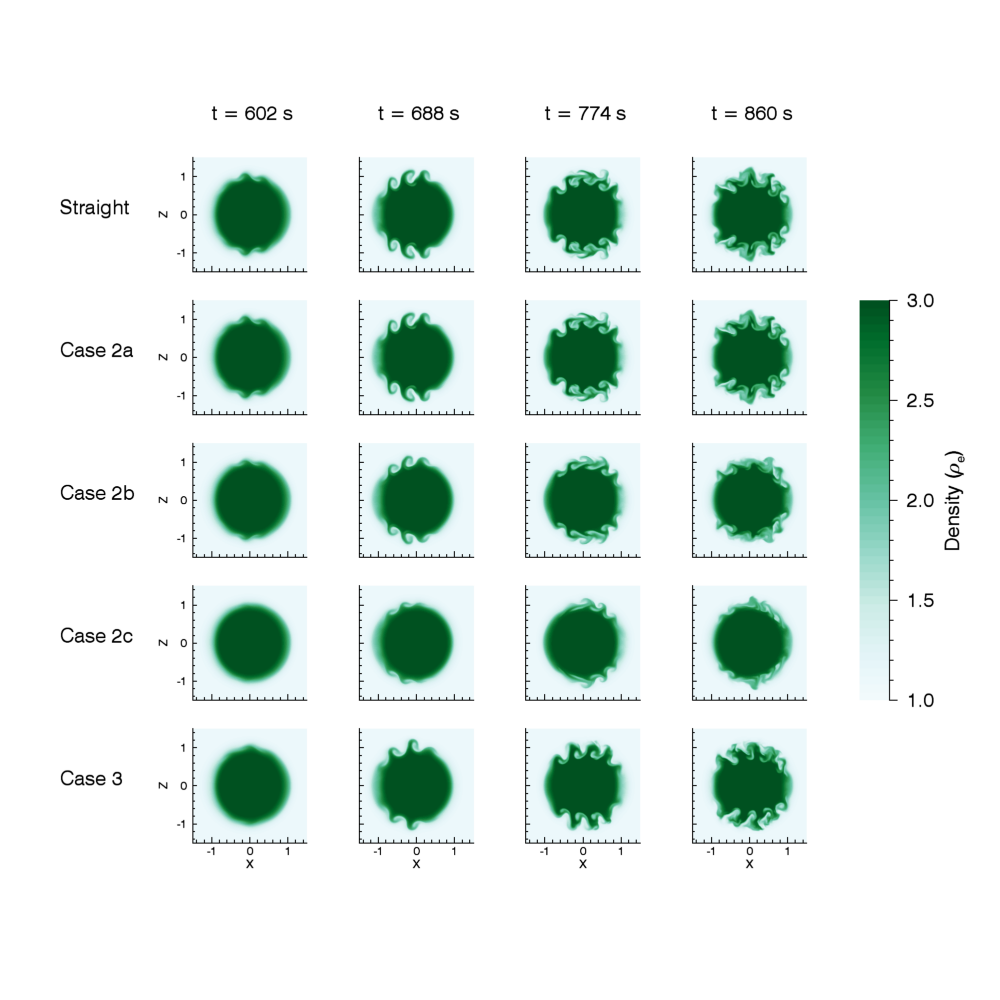}
  \caption{Density deformation at the loop apex during the formation of the KHI. The density is normalised to the initial exterior density $\rho_e$.}
\label{Dens_Grid}
\end{figure*}

In both cases we observe an oscillatory pattern in $v_\phi$ throughout the length of the loop that corresponds to the periodic reversal of the azimuthal waves. The period of these waves match the period of the global kink mode ($\approx 280$ s).  Further, for both the straight and twisted simulations, we observe an increase in the amplitude of these waves through time as the resonant absorption injects energy into the Alfv\'enic modes. In the straight case, we notice that once the initial velocity variation has been accounted for, the energy transfer is roughly constant with height. However, the twisted case displays greater variation along the loop axis and, whilst the model is different, shows evidence of the helical fluting modes found in, for example, \citet{Ruderman2010}. We note that whilst the vertical uniformity is violated in the twisted case, some symmetry is still present within the loop. In particular, if the plot is produced for the line $x=c_x, z=c_z + \zeta$, then we reproduce the lower panel of Figure \ref{Harmonics} except with a reflection in $y=0$ and a change of sign for $v_\phi$.

\subsection{Subsequent dynamics and the growth of the KHI}
The large velocity shear that develops within the boundary region of the loop (see Figure \ref{Res_abs}), becomes unstable to the KHI. The instability is readily observed in the density profile at all heights in the loop, but due to the greatest initial velocity, it is most apparent at the loop apex. In Figure \ref{Dens_Grid} we show the density cross-section at four different stages in the development of the KHI for all of the numerical simulations. 

With the exception of Case 2c, in all of the experiments, Kelvin-Helmholtz vortices begin to form at around $t=600$ s ($ t \approx 680$ s for Case 2c). For a while, three coherent vortices are observed on both sides of the loop \citep[$m=3$, see Paper 1 and eq. 1 in][]{Terradas2008}. Beyond this time, further vortices develop as the instability progresses and ultimately a turbulent regime with small spatial scales is generated. Prior to the onset of the KHI, the twisted cases exhibit only minor differences to the straight case. In particular, these differences would be extremely difficult to detect observationally. However, during the growth of the instability, the azimuthal field begins to have a more profound effect on the evolution of the loop. 

Despite this, in terms of the density deformation, Case 2a displays almost identical behaviour to that of the straight case which may be expected due to the extremely weak twist present $\left(\frac{|B_\phi |}{|B_y|} \le10^{-3} \right)$ within the domain. Increasing the size of the azimuthal component of the magnetic field clearly suppresses the vortex formation. The cause of this inhibition is the magnetic tension force \citep{Soler2010}. Since resonant absorption proceeds in a similar fashion at all heights, the KHI tends to act in an homogeneous manner along the length of the loop. Certainly, observing the density cross-section a few grid points up (in the $y$-direction), looks similar to the panels shown in Figure \ref{Dens_Grid}. Since the magnetic field is approximately frozen into the plasma, this ensures that straight field lines remain locally straight. On the other hand, since the density deformation is not azimuthally constant, twisted field lines experience different KHI effects at different heights. This causes field lines to bend and experience a tension force that acts to suppress further development of the instability.

In addition we observe that the location of twist within the loop can affect the growth of the instability. In particular, Case 3 has the same maximum twist as Case 2c but it is located further from the loop centre. Indeed, this case has a higher quantity of twist when integrated across the entire loop. Despite this, we see a lower level of vortex suppression in the Case 3 simulation than in Case 2c. However, once the first KH vortices have formed, there is evidence that the greater radial extent of the twist restricts the development of a turbulent regime at the edge of the shell region (compare the small scales observed in Row 1 Column 3 with those in Row 5 Column 3) . Whilst the vortices that extend into the region of high twist experience an enhanced tension force that arrests the formation of smaller structures, there is little suppression of turbulence closer to the core of the loop.

\begin{figure}[h]
  \centering
  \includegraphics[width=0.45\textwidth]{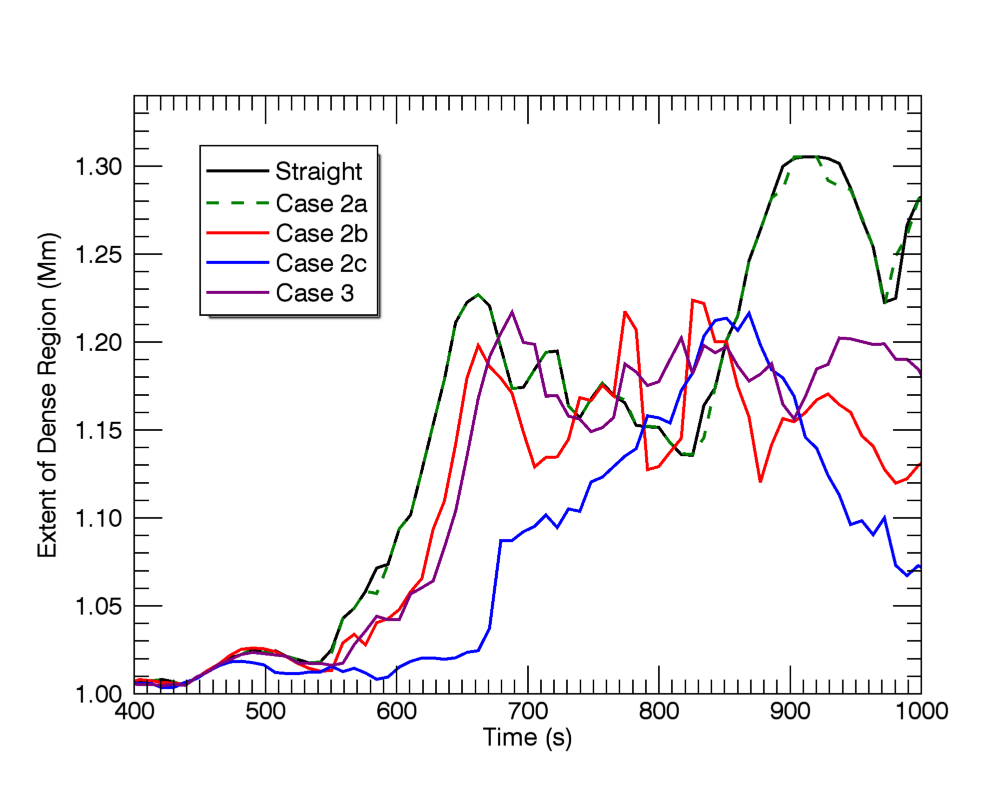}
  \caption{Greatest distance of plama with density, $\rho =2$ from the centre of the loop apex during the formation of the KHI. In the straight field case, initially this plasma is located on the resonance layer.}
\label{Dens_Growth}
\end{figure}

\begin{figure}[h]
  \centering
  \includegraphics[width=0.45\textwidth]{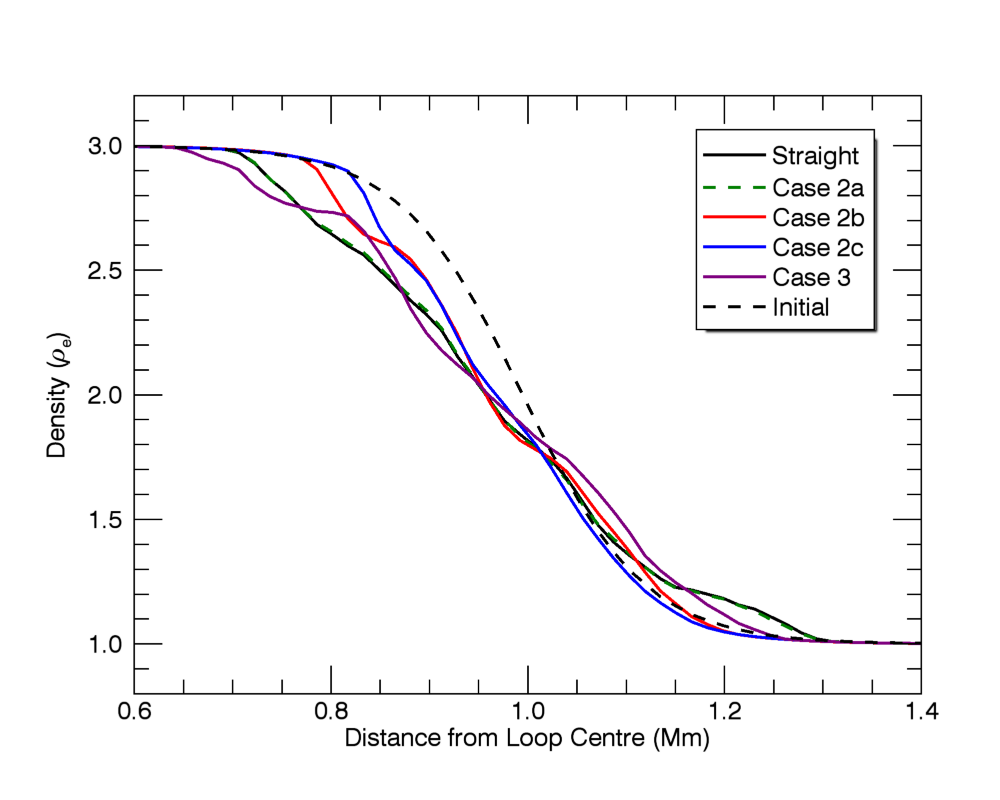}
  \caption{Mean density along loop radii in the horizontal cross-section at the loop apex at $t = 990$ s. By this time, the KHI is well-developed in all cases. For comparison, the initial density profile is also included (black dashed line).}
\label{Dens_Deform}
\end{figure}

In Paper 1, we established several metrics for determining the onset time of the KHI and measuring its subsequent growth. In order to quantify the extent of suppression caused by the magnetic twist, we examine these in Figures \ref{Dens_Growth}-\ref{Dens_Deform}. A further comparison of the vorticity growth that was considered in Paper 1 is addressed in the Vorticity section below. In each figure we observe that the twist either postpones the onset of the KHI (Case 2c) or inhibits its subsequent growth. In the case of weak twist, it is only once the density deformation begins that the suppressive effects of the tension force arise.

In Figure \ref{Dens_Growth}, we aim to summarise the panels of Figure \ref{Dens_Grid}. Tracking the greatest distance of plasma of density, $\rho = 2$ (this corresponds to the location of the resonant layer in the straight case) from the centre of the loop allows the magnitude of the density deformation to be quantified. The suppressive effects of the magnetic twist are clear and again we note that Case 3 exhibits much lower suppression than Case 2c. Further, since the plasma being tracked is at the location of the resonant layer, we expect that any subsequent energy transfer into the Alfv\'enic modes will become less azimuthally uniform \citep{Ruderman2010}.

In Paper 1, we observed that the KHI reduced the mean radial density gradient across the loop boundary (i.e. smoothed the density transition). In particular, simulations in which the KHI was partially or totally suppressed displayed a density profile which remained very similar to the initial density profile. In Figure \ref{Dens_Deform}, we again observe this as the Case 2 simulations exhibit successively less smoothing as the twist was increased.

\begin{figure}[h]
  \centering
  \includegraphics[width=0.5\textwidth]{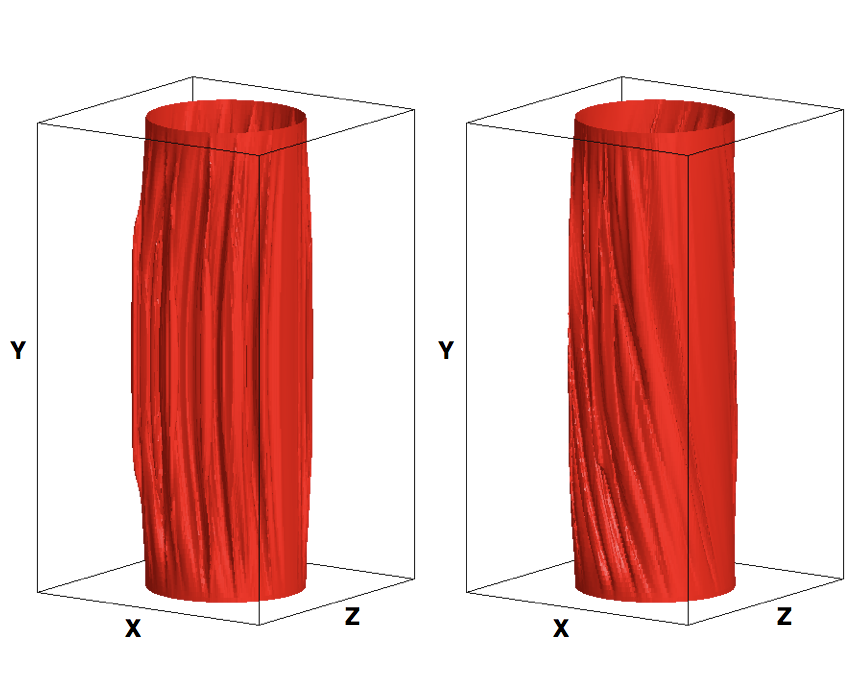}
  \caption{Isosurface of density, $\rho=2$ for the straight field case (left-hand panel) and Case 2c, a twisted field simulation (right-hand panel). The time shown is 990 s (three and a half wave periods) after the start of the simulation when the KHI is well-developed.}
\label{Dens_Comparison}
\end{figure}

In Figure \ref{Dens_Comparison}, we highlight the density deformation caused by the KHI along the length of the loop. In the straight field case (left), the vortices observed are similar at all heights: since the velocity amplitude is not constant along the loop axis, the magnitude of the density deformation is largest at the antinode (loop apex), however, since the velocity shear has the same form (it is proportional to the maximum initial speed at that height), the KH-vortices are similar along the length of the loop. In contrast, the twisted magnetic field case (right) displays helical density deformation that is much smaller in magnitude than that observed in the straight case. This is in good agreement with Figure \ref{Dens_Grid}. 

\begin{figure}[h]
  \centering
  \includegraphics[width=0.5\textwidth]{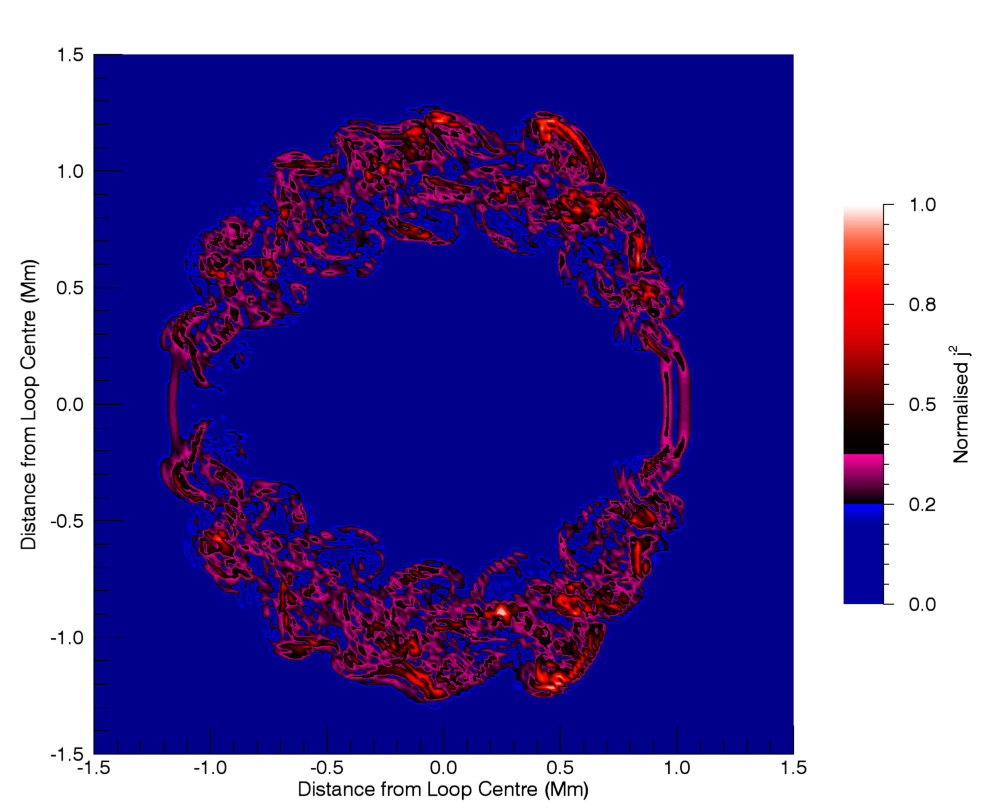}
  \caption{Magnitude of the current in the horizontal plane at the loop apex for the straight field case. Currents less than a fifth of the maximum size of the current in this plane are coloured in blue. By this stage in the simulation ($ t = 990$ s), the KHI is well-developed.}
\label{Cur_apex}
\end{figure}

\begin{figure*}
  \centering
  \includegraphics[width=\textwidth]{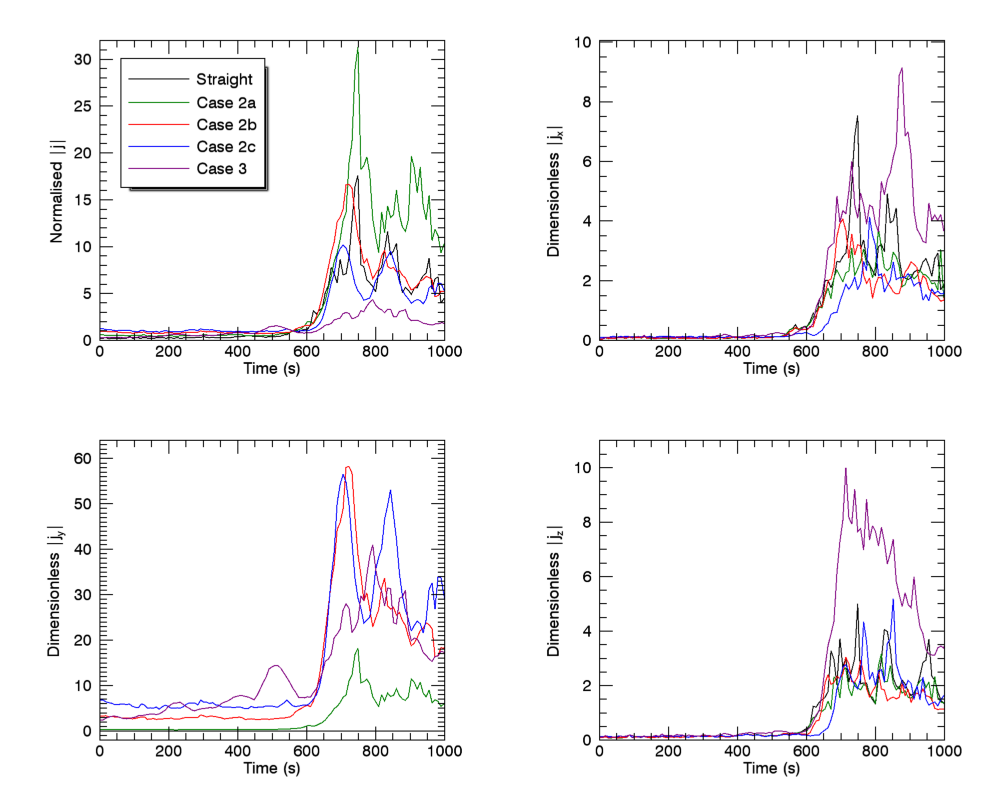}
  \caption{Formation of large currents during the onset of the KHI. The top-left panel displays the maximum magnitude of the current in the horizontal cross-section at the loop apex. For each simulation, it is normalised to the maximum current in this plane at $t = 560$ s, immediately prior to the onset of the KHI, to show the relative growth of the currents during the development of the instability. The other three panels display the maximum magnitude of each component in the same cross-section. The units are dimensionless but are the same for each of the components.}
\label{Current_growth}
\end{figure*}

\subsection{Currents and field-aligned flows}

\begin{figure*}
  \centering
  \includegraphics[width=0.9\textwidth]{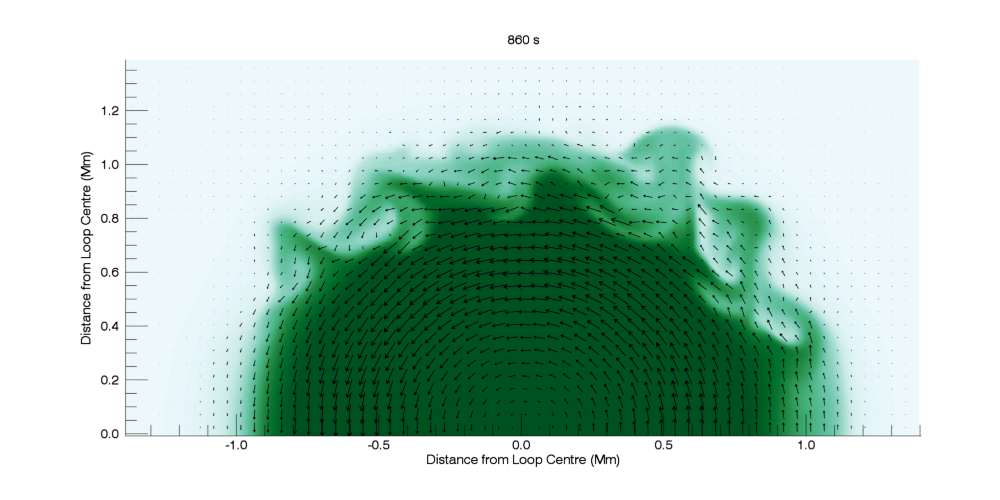}
  \caption{Horizontal components of the magnetic field (vectors) and the deformed density profile (coloured contour plot) at the loop apex once the KHI is well-developed. The result shown is produced from Case 2b, a twisted field simulation.}
\label{Dens_vec}
\end{figure*}

One major effect of including magnetic twist within the oscillating loop is the formation of larger currents. In the model set-up, the azimuthal component of the magnetic field generates a field-aligned current which is in turn associated with a radially inwards tension force. This is initially balanced by adjusting the gas pressure accordingly. During the experiments, the resonant absorption process results in the formation of small scale currents within the loop's boundary layer. In non-ideal simulations these are subject to dissipation which can heat the plasma \citep[e.g.][or see Paper 1]{Pagano2017}. Subsequently, the onset of the KHI generates even smaller scales in the magnetic field and hence larger currents, encouraging further heating (when larger dissipation coefficients are included). 

In Figure \ref{Cur_apex}, we display the magnitude of the current in the loop apex (this is related to the Ohmic heating rate, $\eta j^2$) once the KHI is well-developed (at $t= 990$ s). Intricate currents form throughout the boundary layer of the loop as the magnetic field is distorted by the Kelvin-Helmholtz vortices. The small length scales in the magnetic field, and hence the current, are generated by the compressive flows caused by the instability \citep{KHI_Reconnection}. As such, the vortical structures produced (see Figure \ref{Dens_Grid}) can also be observed in Figure \ref{Cur_apex}. Any Ohmic heating associated with these currents will be confined to the boundary region where the turbulent flows form in these simulations. However, during the development of the KHI in high amplitude regimes \citep[see][]{Magyar2016}, the loop can deform sufficiently to heat the core of the loop through mixing or directly through Ohmic heating as currents begin to develop throughout the cross-section of the loop. Consequently, unlike with resonant absorption induced phase mixing, the core of the loop can be heated in this manner. 

The currents at the loop apex are dominated by the horizontal components $j_x$ and $j_z$ for the straight field case. However, this is not true for the twisted field cases. In Figure \ref{Current_growth}, we display the maximum magnitude of the current (and each component) at the loop apex for each case as the simulations evolve. The currents in the top left-hand panel are normalised to the pre-KHI formation currents in each simulation to allow comparison of current growth factors. From the bottom left-hand panel, it is clear that the twisted cases generate much larger loop-aligned currents than the straight magnetic field case. Meanwhile, with the exception of $|j_z|$ in Case 3 (discussed below), the horizontal currents remain roughly similar in all cases.

Since $\mu \vec{j} = \nabla \times \vec{B}$, the $y$-component of the current is associated with radial gradients in $B_\phi$ and azimuthal gradients in $B_R$. In the straight case, at the loop apex, there is no horizontal field component (with the exception of small variations induced by the KHI) and so $|j_y|$ is small. In the twisted cases, on the other hand, $B_\phi (R)$ is not constant and so there is a loop-aligned component present even prior to the development of the instability ($t < 600$ s; Figure \ref{Current_growth}, bottom left-hand panel). Since the size of the radial gradient is dependent on the magnitude of the twist factor, we see that, for example, Case 2c has larger currents than Case 2b. 

Following the onset of the KHI, the difference is enhanced by the vortical flows. In Figure \ref{Dens_vec}, we display the horizontal components of the magnetic field along with the deformed density profile for Case 2b. Considering a plasma element that is moving radially outwards on account of the KHI, we see that it advects field with a large azimuthal component into a region where there is lower twist. This process decreases the length scales on which the azimuthal component varies and so increases the currents. Further twisting motions associated with the KHI introduce radial components of the field and generates further currents.   

Despite the suppressive effects of the azimuthal magnetic field on the density deformation, the enhanced currents observed will result in greater KHI-related heating for twisted regimes through Ohmic dissipation. Indeed, the third panel of Figure \ref{Current_growth} does demonstrate that the KHI in twisted cases is more energetic than in the straight case. However, as we see by comparing the Case 2 simulations in the top left hand panel of Figure \ref{Current_growth}, increasing the strength of the magnetic twist produces smaller relative increases (from the pre-KHI level) in the magnitude of the current. We also note that despite the very small amount of magnetic twist present in Case 2a, there are still significant loop-aligned currents generated. This suggests that even though the field in Case 2a is locally almost straight, it may produce different heating signatures than the truly straight field case. 

As mentioned above, Case 3 produces currents with a larger component in the $z$ direction than the other cases. A later peak is also observed in $j_x$ for Case 3 (top right-hand panel) which is larger than the $x$ components of the currents generated in the other simulations. The horizontal components of the current reach much larger values in this case because the KH-vortices form in regions of stronger magnetic twist than in the Case 2 simulations. The transverse gradients produced by the KHI are therefore much greater than those that form in the other simulations.

One further consequence of the magnetic twist is the excitation of loop-aligned velocities. At all times in the straight case, any vertical flows present at the loop apex are extremely small. Since the gas and magnetic pressures are vertically uniform, there is no $y$-component of the magnetic tension and gravity is neglected, there are no forces to drive vertical flows at the loop apex. However, once twist in the magnetic field is introduced, the radial currents that form during the growth of the KHI, produce a vertical component of the Lorentz force. This is the case even for small amounts of magnetic field twist. Since these currents are highly spatially variable (some are orientated radially outwards and some are orientated radially inwards), both upwards and downward flows are generated.

\begin{figure}[h]
  \centering
  \includegraphics[width=0.5\textwidth]{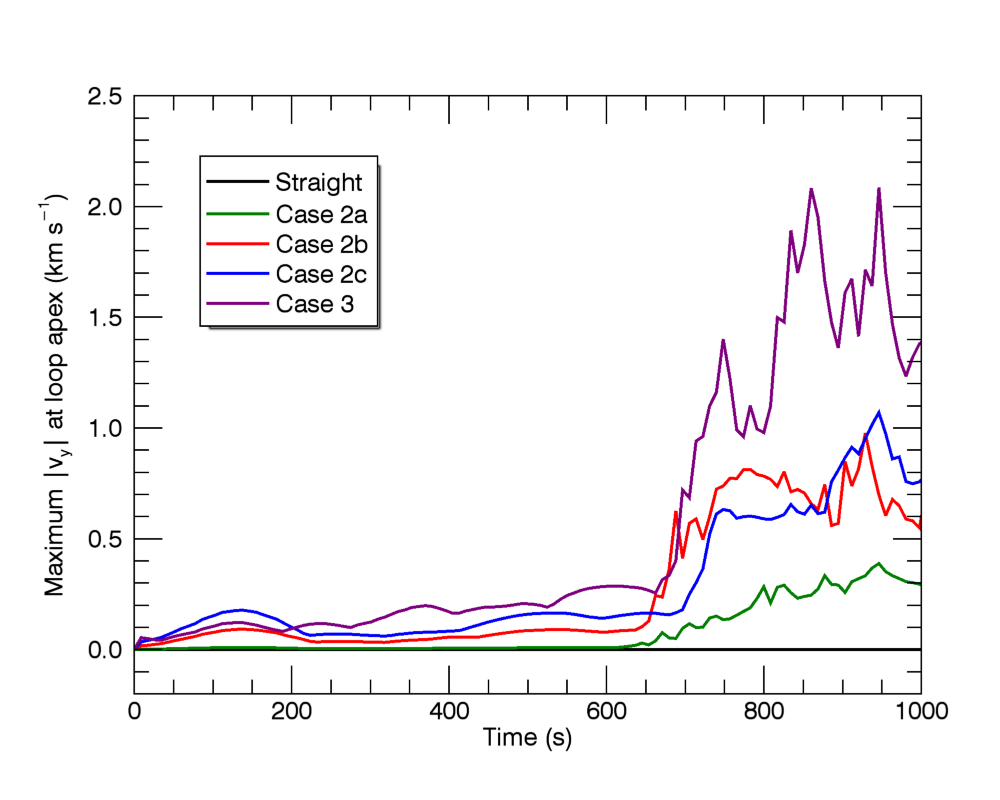}
  \caption{Maximum value of $\left|v_y\right|$ at the loop apex against time for all simulations.}
\label{Vy_growth}
\end{figure}

In Figure \ref{Vy_growth}, we show the maximum value of $v_y$ at the loop apex for all the simulations. We note that although Case 2a does not seem too different from the straight case, the velocities observed in the weakly twisted case are many orders of magnitude larger than those in the straight case. In all of the twisted cases, an increase in vertical flows is observed during the resonant absorption phase ($ t < 600 $ s), as currents forming in the shell region interact with the azimuthal magnetic field, generating a vertical Lorentz force. This effect becomes much more apparent as the KHI forms and larger currents are generated. 

Case 3 displays significantly larger vertical flows then the Case 2 simulations due to the magnetic twist, and hence vertical Lorentz forces, being larger in the shell region where the largest currents form. Additionally, since the horizontal currents in the loop apex are greater in Case 3 (see upper right-hand panel of Figure \ref{Current_growth}, this effect is further enhanced. Even prior to the formation of the KHI, we observe some loop-aligned flows being produced due to the interaction of the azimuthal magnetic field and the currents associated with the resonant absorption. Once the KHI forms, the suppressive effect of the magnetic twist ensures that for a short period the vertical flows are greater in Case 2b than Case 2c despite the stronger twist component present in the latter simulation.

\subsection{Vorticity}

\begin{figure}
  \centering
  \includegraphics[width=0.45\textwidth]{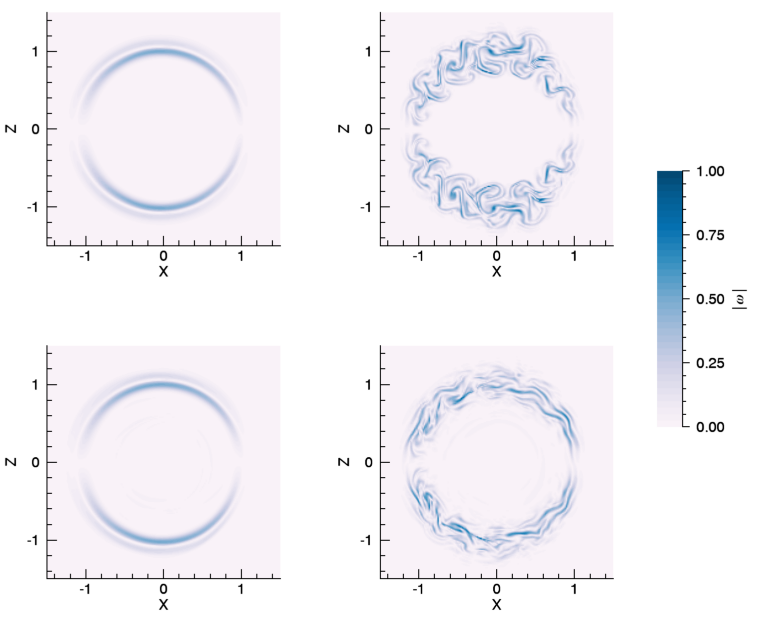}
  \caption{Magnitude of the vorticity at two different times, $ t = 420 $ (left) and $t = 990$ s (right) in the horizontal cross-section through the loop apex. The top row corresponds to Case 1, the straight field simulation, and the second row corresponds to Case 2c. Here we have normalised the vorticity to the maximum observed in either simulation.}
  \label{Vortgrid}
\end{figure}

\begin{figure}
  \centering
  \includegraphics[width=0.45\textwidth]{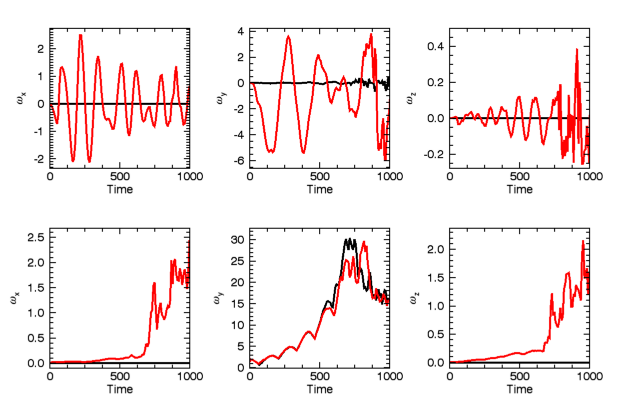}
  \caption{Top row - the three components of the vorticity integrated over the horizontal plane at the loop apex. Bottom row - maximum size of each component of the vorticity in this plane. For both of the rows, the black line corresponds to the straight field simulation (Case 1) and the red line corresponds to Case 2c.}
\label{Vortosc}
\end{figure}

In addition to the small scales that form in the magnetic field during the development of the KHI, small scales in the velocity field are also generated (e.g. Paper 1). In non-ideal simulations, these gradients are susceptible to the effects of viscosity which, in turn, lead to the dissipation of wave energy. During resonant absorption, the excitation of azimuthal Alfv\'enic waves is associated with an increase in the loop-aligned component of vorticity. Following the subsequent growth of the KH vortices, the magnitude of the vorticity increases further.

In Figure \ref{Vortgrid}, we display the magnitude of the vorticity, $|\omega|$ in the horizontal cross-section at the loop apex. We show two different stages during the simulation for the straight field case (top row) and the twisted field Case 2c (bottom row). Prior to the formation of the instability, the profiles of $|\omega|$ are very similar. Once the KHI forms, differences become apparent as the characteristic vortices are suppressed in the twisted field case. However, the maximum size of the vorticity is very similar in both of these simulations, suggesting that despite the limited spatial extent of the vortices, the amount of kinetic energy is similar.

In the top row of Figure \ref{Vortosc}, we display the three components of the vorticity integrated over the loop cross-section. In particular, at each time we calculate,

\begin{equation} \int_A \omega_x \, \text{d} A = \int_A \frac{\partial v_z}{\partial y}  - \frac{\partial v_y}{\partial z} \, \text{d} A, \end{equation}

\begin{equation} \int_A \omega_y \, \text{d} A = \int_A \frac{\partial v_x}{\partial z}  - \frac{\partial v_z}{\partial x} \, \text{d} A, \end{equation}

\begin{equation} \int_A \omega_z \, \text{d} A = \int_A \frac{\partial v_y}{\partial x}  - \frac{\partial v_x}{\partial y} \, \text{d} A, \end{equation}

for the panels from left to right. Here the area $A$ is the horizontal cross-section at the loop apex. 
For all three components of $\omega$, the straight field simulation (black lines) displays very little net vorticity when integrated across the loop cross-section. On the other hand, the inclusion of magnetic twist permits the non-linear generation of additional harmonics that are associated with the oscillating profiles in the panels in the top row of Figure \ref{Vortosc}.  

The most significant effect is observed in $\omega_y$ (second column). In the straight case, the azimuthal waves in the upper $z$ half-plane have an equal and opposite contribution to those in the lower $z$ half-plane. Meanwhile in the twisted case, the magnetic tension force associated with the azimuthal field has a positive contribution to one of the half-planes and has an inhibiting effect in the other. This causes the $\omega_y$ contribution of one set of azimuthal waves to be slightly larger than the set in the opposite half of the domain. The effect of the tension force reverses as the direction of motion changes, thus causing the observed oscillatory behaviour.

For the other two components of vorticity, the loop integrated values oscillate with twice the frequency of the kink mode. For both $\omega_x$ and $\omega_z$, the associated panels in Figure \ref{Vortosc} are dominated by the contribution of loop-aligned gradients; $\frac{\partial v_z}{\partial y}$ and $\frac{\partial v_x}{\partial y}$, respectively. These gradients are associated with additional, higher harmonic azimuthal waves that are generated in the twisted field cases but are not observed in the straight field case.

Finally, the bottom row of Figure \ref{Vortosc} displays the maximum size of each component of the vorticity in the same horizontal plane. The central panel was used as a proxy for the growth of the KHI in Paper 1. We see that despite being slightly suppressed when the KHI first forms, the maximum vorticity in the twisted case exceeds that observed in the straight case at arount $t= 800$ s. This supports the claim that the vortices in Case 2c have as much kinetic energy as the straight field vortices. There is a marked difference in the growth of the other two components of vorticity. In particular, the maximum size of $|\omega_x|$ and $|\omega_z|$ is around an order of magnitude larger in the twisted field case. This is caused by the azimuthal waves and the KH vortices having a significant helical component (see Figure \ref{Dens_Comparison}).

\subsection{Loop-aligned numerical resolution}
As shown above, one result of the inclusion of magnetic twist is that the K-H vortices are helical in nature. Indeed, vertical structure in the density and velocity field becomes more profound in Cases 2 and 3 and, as more twist is included, this effect increases. Hence, both the horizontal and vertical resolution might affect the development of the KHI. Since, in our simulations, the loop-aligned spatial resolution is much coarser than the horizontal spatial resolution, we here investigate whether the vertical resolution is sufficient.

In all cases, the most unstable wave vector satisfies $\vec{k} \cdot \vec{B} = 0$ such that the inhibiting effect of the right-hand side of equation 4.7 in \citet{Cowling1976} is eliminated. Thus, an additional consequence of the inclusion of magnetic twist in our model is the modification of this wave vector, $\vec{k}$. The azimuthal component of the magnetic field in Cases 2 and 3 ensures that $\vec{k}$ has a non-zero component parallel to the loop axis. In particular, it is no longer confined to the $x$-$z$-plane as it is in the straight field case.

We estimate the required vertical resolution as follows. Since $\vec{k} \cdot \vec{B} = 0$ and $B_R \approx 0$ when the instability is triggered, we have

\begin{equation} \label{Wavelength1} \frac{\left|k_y\right|} {\left|k_\phi\right|} \approx \frac{\left|B_\phi\right|}{\left|B_y\right|}.\end{equation}

Meanwhile, in order to resolve the vertical component of $\vec{k}$ at least as well as we resolve the horizontal component, we require

\begin{equation} \label{Wavelength2} \Delta_y \le  \frac{\lambda_{y}\Delta_{\phi}}{\lambda_{\phi}},   \end{equation}

where $\Delta_y$ and $\Delta_{\phi}$ are the vertical and azimuthal numerical resolution (grid widths) and $\lambda_y = \frac{2 \pi}{k_y}$ and $\lambda_{\phi} =  \frac{2 \pi}{k_{\phi}}$ are the vertical and azimuthal wave lengths, respectively. We note that $\Delta_{\phi}$ is confined by the horizontal grid widths $\Delta_x$ and $\Delta_z$ such that 

\begin{equation}\Delta_x \le \Delta_{\phi} \le \sqrt{\Delta_x^2 + \Delta_z^2}.\end{equation}

Consequently, since equation (\ref{Wavelength2}) must be satisfied throughout the instability forming region, we obtain

\begin{equation} \label{upperbound} \Delta_y \le \frac{B_y \Delta_x}{B_{\phi}}. \end{equation}

In particular, we immediately notice that the greater the magnetic twist, $B_{\phi}$, the more refined the vertical grid has to be in order to ensure the simulation is well resolved. We also note this is only an upper bound on $\Delta_y$, and in the limit $B_{\phi} \to 0$, more restrictive bounds are required in order to resolve the initial kink mode, for example. 

In our model, for the most twisted cases, equation (\ref{upperbound}) yields $\Delta_y \lesssim 1200$ km, suggesting that the vertical component of the wave vector $\vec{k}$ is less well resolved than the horizontal component. In order to determine whether this produced significant numerical artefacts in our simulations, we re-used the set-up of Case 2c but with double the number of grid points in the $y$ direction.

\begin{figure}
  \centering
  \includegraphics[width=0.45\textwidth]{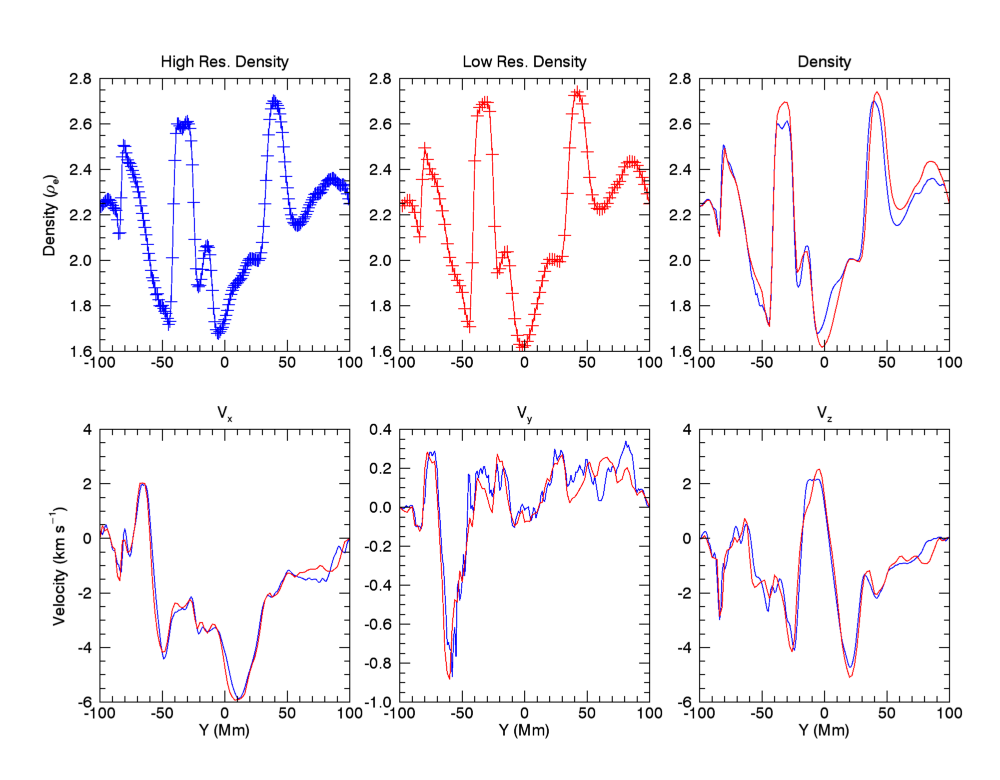}
  \caption{Top row - density profile along the most under-resolved vertical line at $t = 900$ s for the high resolution (blue) and low resolution (red) simulations. In the first two panels, the position of grid points are denoted with a cross, and the third panel shows a comparison between the two simulations.  Bottom row - comparison of the three velocity components along the same line for the two different resolutions.}
\label{Res_compare}
\end{figure}

In Figure \ref{Res_compare}, we display the density and velocity profiles along the most under-resolved vertical line in the low resolution simulation at a time when the instability is well-developed. The line lies within the vortex-forming region in the boundary of the loop. We observe that for -45 Mm $\le y \le$ -35 Mm, the density profile is approaching the limit of the spatial resolution, particularly in the low resolution (100 grid points) simulation. Despite this, there is little difference between the density profile (third panel) and the velocity field (bottom row) produced by the two experiments. We found that there was no significant difference in the growth rate of the K-H vortices or indeed in any of the plots displayed previously. However, we note that these resolution effects are expected to become more significant in regimes with greater magnetic twist.

\section{Discussion and conclusions}
We have conducted an investigation into the effects of magnetic twist on the dynamics of a transversely oscillating coronal loop. In all cases considered, the level of twist was weak in the sense that $B_\phi \ll B_y$ (where $B_y$ is the loop-aligned component of the field) and the loop was not kink unstable.  

The fundamental standing kink wave was induced by an initial velocity profile and was observed to decay due to resonant absorption. Initially, this phenomenon, through which energy is transferred from the standing kink mode to localised Alfv\'enic waves, proceeds largely independently of the twist present within the loop. The azimuthal component of the magnetic field does modify the Alfv\'enic modes slightly, but the global dynamics remain unchanged when compared to the straight field case (see Paper 1). In particular, the twisted field permits an additional, higher harmonic, azimuthal wave that contributes an oscillating net vorticity at the loop apex.

The azimuthal waves are contained within the boundary region of the loop and are associated with a significant radial velocity shear. In dissipative media, these waves will deposit their energy as heat through the process of phase mixing; however, due to the approximately ideal regime presented, this was not observed. Once a critical shear has been reached, the Kelvin-Helmholtz instability is triggered in all cases. The density deformation that is associated with the KHI is partially suppressed in cases with twisted magnetic field. Further, for the extent of the parameter space considered here, more twist causes greater suppression. In contrast with the dissipation dependent suppression observed in Paper 1, we note that the twist does not significantly delay the onset of the KHI, but instead limits its subsequent growth. This is in agreement with the velocity shear associated with the resonant absorption growing independently of the twist.

Previous studies have demonstrated that other parameters such as the amplitude of the initial perturbation, the width of the boundary layer \citep{Magyar2016, Terradas2017}, and the dissipation coefficients \citep{Howson2017} can modify the growth rate of the instability. In the first of these articles, the authors find that decreasing the amplitude of the initial perturbation decreases the instability's growth rate. Similarly, the second work finds that in both straight and twisted field regimes, thicker boundary layers restrict the growth rate of the KHI.  Both of these effects, along with the stabilisation via enhanced dissipation coefficients, are caused by a reduced velocity shear associated with the Alfv\'en waves. In the case of twisted field, however, the shear is only weakly affected and instead the instability growth rate is inhibited by the magnetic tension force.

Despite the suppression of the density deformation, the KHI in twisted fields may have significance for the coronal heating problem because it generates the small length scales required for an enhanced dissipation rate \citep{Antolin2014, Howson2017}. Indeed, Paper 1 demonstrated that even with lower dissipation coefficients, the instability may result in earlier heating than is produced by phase mixing alone at high dissipation. Whilst the suppression of the KH vortices is readily observed in both the density cross-section and the horizontal velocity field, smaller spatial scales form in the magnetic field in the cases with greater twist. Since twisted cases have an azimuthal component that is radially non-uniform, the characteristic, radial movement of plasma, and hence magnetic field (since it is approximately frozen-in to the plasma), associated with the KHI can generate large currents. Indeed, if coronal loops are twisted in some way, then some of the density deformation caused by the KHI would be suppressed, but importantly stronger currents are generated and hence greater Ohmic heating is expected. Meanwhile, analysis of the vorticity at the loop apex reveals that although the density deformations produced in the twisted cases are smaller in size, they have as much kinetic energy as in the straight case.

Additionally, once the KHI currents begin to form, the azimuthal component of the magnetic field produces a vertical component of the Lorentz force that drives upward and downward flows. These are not observed in the completely straight case but are significant even in weakly twisted regimes. This offers the possibility of detecting magnetic twist in coronal loops by searching for loop-aligned flows at the apex in decaying kink oscillations. 

One further effect of the inclusion of magnetic twist is that, contrary to the straight case, the development of the KHI triggered vertical asymmetry along the loop (as a consequence of the loop-aligned flows). This was observed both in the velocity field and the cross-sectional density profile. From a regime that was vertically symmetric about the loop apex, the instability generated significant non-uniformity between the two halves of the loop. When magnetic twist is included, these effects should be considered when vertical symmetry is used in order to model only one half of the loop and hence save computational time. 

A major criticism of previous work considering the Kelvin-Helmholtz instability as a driver of MHD turbulence in the corona and hence a possible coronal heating mechanism is that even weak magnetic twist will suppress the development of the instability. Whilst we have demonstrated that, to some extent, this is indeed the case, the growth of much larger currents in the twisted field cases more than compensates for the level of suppression when considering energy dissipation. The ultimate effect of the twist is an increase in the difficulty of detecting the KHI vortices observationally rather than a decrease in its efficiency as a heating mechanism. Whilst the reduced density deformation may reduce the chance of observing the instability directly, the presence of field aligned flows may offer an alternative detection mechanism.

\vspace{1cm}

The authors would like to thank the referee for their helpful comments and suggestions. The authors would also like to thank Prof. P. J. Cargill for helpful discussions. The research leading to these results has received funding from the UK Science and Technology Facilities Council (consolidated grant ST/N000609/1) and the European Union Horizon 2020 research and innovation programme (grant agreement No. 647214). This work used the DIRAC 1, UKMHD Consortium machine at the University of St Andrews and the Darwin Data Analytic system at the University of Cambridge, operated by the University of Cambridge High Performance Computing Service on behalf of the STFC DiRAC HPC Facility (www.dirac.ac.uk). This equipment was funded by a BIS National E-infrastructure capital grant (ST/K001590/1), STFC capital grants ST/H008861/1 and ST/H00887X/1, and DiRAC Operations grant ST/K00333X/1. DiRAC is part of the National E-Infrastructure.

\bibliographystyle{aa}        
\bibliography{KHI_Twist}           

\begin{thebibliography}{53}
\expandafter\ifx\csname natexlab\endcsname\relax\def\natexlab#1{#1}\fi

\bibitem[{{Andries} {et~al.}(2005){Andries}, {Goossens}, {Hollweg}, {Arregui},
  \& {Van Doorsselaere}}]{Andries2005}
{Andries}, J., {Goossens}, M., {Hollweg}, J.~V., {Arregui}, I., \& {Van
  Doorsselaere}, T. 2005, \aap, 430, 1109

\bibitem[{{Anfinogentov} {et~al.}(2015){Anfinogentov}, {Nakariakov}, \&
  {Nistic{\`o}}}]{Anfino2015}
{Anfinogentov}, S.~A., {Nakariakov}, V.~M., \& {Nistic{\`o}}, G. 2015, \aap,
  583, A136

\bibitem[{{Antolin} {et~al.}(2015){Antolin}, {Okamoto}, {De Pontieu},
  {Uitenbroek}, {Van Doorsselaere}, \& {Yokoyama}}]{Antolin2015}
{Antolin}, P., {Okamoto}, T.~J., {De Pontieu}, B., {et~al.} 2015, \apj, 809, 72

\bibitem[{{Antolin} {et~al.}(2014){Antolin}, {Yokoyama}, \& {Van
  Doorsselaere}}]{Antolin2014}
{Antolin}, P., {Yokoyama}, T., \& {Van Doorsselaere}, T. 2014, Astrophys. J.\
  Letts., 787, L22

\bibitem[{{Arber} {et~al.}(2001){Arber}, {Longbottom}, {Gerrard}, \&
  {Milne}}]{Larey}
{Arber}, T.~D., {Longbottom}, A.~W., {Gerrard}, C.~L., \& {Milne}, A.~M. 2001,
  Journal of Computational Physics, 171, 151

\bibitem[{{Arregui}(2015)}]{Arregui2015}
{Arregui}, I. 2015, Philosophical Transactions of the Royal Society of London
  Series A, 373, 20140261

\bibitem[{{Aschwanden} {et~al.}(2002){Aschwanden}, {de Pontieu}, {Schrijver},
  \& {Title}}]{Aschwanden2002}
{Aschwanden}, M.~J., {de Pontieu}, B., {Schrijver}, C.~J., \& {Title}, A.~M.
  2002, \solphys, 206, 99

\bibitem[{{Aschwanden} {et~al.}(1999){Aschwanden}, {Fletcher}, {Schrijver}, \&
  {Alexander}}]{Aschwanden1999}
{Aschwanden}, M.~J., {Fletcher}, L., {Schrijver}, C.~J., \& {Alexander}, D.
  1999, \apj, 520, 880

\bibitem[{{Browning} \& {Priest}(1984)}]{Browning1984}
{Browning}, P.~K. \& {Priest}, E.~R. 1984, \aap, 131, 283

\bibitem[{{Cargill} {et~al.}(2016){Cargill}, {De Moortel}, \& {Kiddie}}]{IDMPC}
{Cargill}, P.~J., {De Moortel}, I., \& {Kiddie}, G. 2016, Astron.\ Astrophys.,
  823, 31

\bibitem[{{Cowling}(1976)}]{Cowling1976}
{Cowling}, T.~G. 1976, {Magnetohydrodynamics} (Adam Hilger, Ltd.)

\bibitem[{{D{\'e}moulin} {et~al.}(2002){D{\'e}moulin}, {Mandrini}, {van
  Driel-Gesztelyi}, {Thompson}, {Plunkett}, {Kov{\'a}ri}, {Aulanier}, \&
  {Young}}]{Demoulin2002}
{D{\'e}moulin}, P., {Mandrini}, C.~H., {van Driel-Gesztelyi}, L., {et~al.}
  2002, \aap, 382, 650

\bibitem[{{Ebrahimi} \& {Karami}(2016)}]{Ebrahimi2016}
{Ebrahimi}, Z. \& {Karami}, K. 2016, \mnras, 462, 1002

\bibitem[{{Fan}(2009)}]{Fan2009}
{Fan}, Y. 2009, \apj, 697, 1529

\bibitem[{{Foullon} {et~al.}(2011){Foullon}, {Verwichte}, {Nakariakov},
  {Nykyri}, \& {Farrugia}}]{Foullon2011}
{Foullon}, C., {Verwichte}, E., {Nakariakov}, V.~M., {Nykyri}, K., \&
  {Farrugia}, C.~J. 2011, Astrophys. J.\ Letts., 729, L8

\bibitem[{{Goossens} \& {Ruderman}(2011{\natexlab{a}})}]{Goossens2011_Rev}
{Goossens}, M.~{Erd\'elyi}, R. \& {Ruderman}, M.~S. 2011{\natexlab{a}}, Space
  Science Reviews, 158, 289

\bibitem[{{Goossens} \& {Ruderman}(2011{\natexlab{b}})}]{Goossens2012}
{Goossens}, M.~{Erd\'elyi}, R. \& {Ruderman}, M.~S. 2011{\natexlab{b}}, Space
  Science Reviews, 158, 289

\bibitem[{{Heyvaerts} \& {Priest}(1983)}]{Heyvaerts1983}
{Heyvaerts}, J. \& {Priest}, E.~R. 1983, Astron.\ Astrophys., 117, 220

\bibitem[{{Hollweg} {et~al.}(1990){Hollweg}, {Yang}, {Cadez}, \&
  {Gakovic}}]{Hollweg1990}
{Hollweg}, J.~V., {Yang}, G., {Cadez}, V.~M., \& {Gakovic}, B. 1990, \apj, 349,
  335

\bibitem[{{Hood} {et~al.}(2013){Hood}, {Ruderman}, {Pascoe}, {De Moortel},
  {Terradas}, \& {Wright}}]{Gauss_Damp1}
{Hood}, A.~W., {Ruderman}, M., {Pascoe}, D.~J., {et~al.} 2013, \aap, 551, A39

\bibitem[{{Howson} {et~al.}(2017){Howson}, {De Moortel}, \&
  {Antolin}}]{Howson2017}
{Howson}, T.~A., {De Moortel}, I., \& {Antolin}, P. 2017, \aap, 602, A74

\bibitem[{{Ionson}(1978)}]{Ionson1978}
{Ionson}, J.~A. 1978, Astrophys. J., 226, 650

\bibitem[{{Karami} \& {Barin}(2009)}]{Karami2009}
{Karami}, K. \& {Barin}, M. 2009, \mnras, 394, 521

\bibitem[{{Lapenta} \& {Knoll}(2003)}]{KHI_Reconnection}
{Lapenta}, G. \& {Knoll}, D.~A. 2003, Solar Physics, 214, 107

\bibitem[{{Magara} \& {Longcope}(2003)}]{Magara2003}
{Magara}, T. \& {Longcope}, D.~W. 2003, \apj, 586, 630

\bibitem[{{Magyar} \& {Van Doorsselaere}(2016)}]{Magyar2016}
{Magyar}, N. \& {Van Doorsselaere}, T. 2016, \aap, 595, A81

\bibitem[{{Morton} {et~al.}(2014){Morton}, {Verth}, {Hillier}, \&
  {Erd{\'e}lyi}}]{Morton2014}
{Morton}, R.~J., {Verth}, G., {Hillier}, A., \& {Erd{\'e}lyi}, R. 2014, \apj,
  784, 29

\bibitem[{{Ofman} \& {Thompson}(2011)}]{Ofman2011}
{Ofman}, L. \& {Thompson}, B.~J. 2011, \apjl, 734, L11

\bibitem[{{Okamoto} {et~al.}(2015){Okamoto}, {Antolin}, {De Pontieu},
  {Uitenbroek}, {Van Doorsselaere}, \& {Yokoyama}}]{Okamoto2015}
{Okamoto}, T.~J., {Antolin}, P., {De Pontieu}, B., {et~al.} 2015, \apj, 809, 71

\bibitem[{{Okamoto} {et~al.}(2007){Okamoto}, {Tsuneta}, {Berger}, {Ichimoto},
  {Katsukawa}, {Lites}, {Nagata}, {Shibata}, {Shimizu}, {Shine}, {Suematsu},
  {Tarbell}, \& {Title}}]{Okamoto2007}
{Okamoto}, T.~J., {Tsuneta}, S., {Berger}, T.~E., {et~al.} 2007, Science, 318,
  1577

\bibitem[{{Pagano} \& {De Moortel}(2017)}]{Pagano2017}
{Pagano}, P. \& {De Moortel}, I. 2017, Astron.\ Astrophys., submitted

\bibitem[{{Parnell} \& {De Moortel}(2012)}]{Parnell2012}
{Parnell}, C.~E. \& {De Moortel}, I. 2012, Philosophical Transactions of the
  Royal Society of London Series A, 370, 3217

\bibitem[{{Pascoe} {et~al.}(2016){Pascoe}, {Goddard}, {Nistic{\`o}},
  {Anfinogentov}, \& {Nakariakov}}]{Pascoe2016}
{Pascoe}, D.~J., {Goddard}, C.~R., {Nistic{\`o}}, G., {Anfinogentov}, S., \&
  {Nakariakov}, V.~M. 2016, Astron.\ Astrophys., 585, L6

\bibitem[{{Pascoe} {et~al.}(2011){Pascoe}, {Wright}, \& {De
  Moortel}}]{Pascoe2011}
{Pascoe}, D.~J., {Wright}, A.~N., \& {De Moortel}, I. 2011, \apj, 731, 73

\bibitem[{{Ruderman}(2003)}]{Ruderman2003}
{Ruderman}, M.~S. 2003, \aap, 409, 287

\bibitem[{{Ruderman} \& {Erd{\'e}lyi}(2009)}]{Ruderman2009}
{Ruderman}, M.~S. \& {Erd{\'e}lyi}, R. 2009, \ssr, 149, 199

\bibitem[{{Ruderman} \& {Goossens}(2014)}]{Ruderman2014}
{Ruderman}, M.~S. \& {Goossens}, M. 2014, \solphys, 289, 1999

\bibitem[{{Ruderman} {et~al.}(2010){Ruderman}, {Goossens}, \&
  {Andries}}]{Ruderman2010}
{Ruderman}, M.~S., {Goossens}, M., \& {Andries}, J. 2010, Physics of Plasmas,
  17, 082108

\bibitem[{{Ruderman} \& {Terradas}(2013)}]{Ruderman2013}
{Ruderman}, M.~S. \& {Terradas}, J. 2013, \aap, 555, A27

\bibitem[{{Rust} \& {Kumar}(1996)}]{Rust1996}
{Rust}, D.~M. \& {Kumar}, A. 1996, \apjl, 464, L199

\bibitem[{{Soler} {et~al.}(2010){Soler}, {Terradas}, {Oliver}, {Ballester}, \&
  {Goossens}}]{Soler2010}
{Soler}, R., {Terradas}, J., {Oliver}, R., {Ballester}, J.~L., \& {Goossens},
  M. 2010, \apj, 712, 875

\bibitem[{{Terradas} {et~al.}(2008{\natexlab{a}}){Terradas}, {Andries},
  {Goossens}, {Arregui}, {Oliver}, \& {Ballester}}]{Terradas2008}
{Terradas}, J., {Andries}, J., {Goossens}, M., {et~al.} 2008{\natexlab{a}},
  \apjl, 687, L115

\bibitem[{{Terradas} {et~al.}(2008{\natexlab{b}}){Terradas}, {Arregui},
  {Oliver}, {Ballester}, {Andries}, \& {Goossens}}]{Terradas2008b}
{Terradas}, J., {Arregui}, I., {Oliver}, R., {et~al.} 2008{\natexlab{b}}, \apj,
  679, 1611

\bibitem[{{Terradas} {et~al.}(2017){Terradas}, {Magyar}, \& {Van
  Doorsselaere}}]{Terradas2017}
{Terradas}, J., {Magyar}, N., \& {Van Doorsselaere}, T. 2017, In preparation

\bibitem[{{Terradas} {et~al.}(2006){Terradas}, {Oliver}, \&
  {Ballester}}]{Terradas2006}
{Terradas}, J., {Oliver}, R., \& {Ballester}, J.~L. 2006, \apjl, 650, L91

\bibitem[{{Tomczyk} {et~al.}(2007){Tomczyk}, {McIntosh}, {Keil}, {Judge},
  {Schad}, {Seeley}, \& {Edmondson}}]{Tomczyk2007}
{Tomczyk}, S., {McIntosh}, S.~W., {Keil}, S.~L., {et~al.} 2007, Science, 317,
  1192

\bibitem[{{T{\"o}r{\"o}k} \& {Kliem}(2003)}]{Torok2003}
{T{\"o}r{\"o}k}, T. \& {Kliem}, B. 2003, \aap, 406, 1043

\bibitem[{{Uchimoto} {et~al.}(1991){Uchimoto}, {Strauss}, \&
  {Lawson}}]{Uchimoto1991}
{Uchimoto}, E., {Strauss}, H.~R., \& {Lawson}, W.~S. 1991, \solphys, 134, 111

\bibitem[{{van Ballegooijen} {et~al.}(2011){van Ballegooijen}, {Asgari-Targhi},
  {Cranmer}, \& {DeLuca}}]{VanBalle2011}
{van Ballegooijen}, A.~A., {Asgari-Targhi}, M., {Cranmer}, S.~R., \& {DeLuca},
  E.~E. 2011, \apj, 736, 3

\bibitem[{{Van Doorsselaere} {et~al.}(2004){Van Doorsselaere}, {Debosscher},
  {Andries}, \& {Poedts}}]{VanDoors2004}
{Van Doorsselaere}, T., {Debosscher}, A., {Andries}, J., \& {Poedts}, S. 2004,
  \aap, 424, 1065

\bibitem[{{Verth} {et~al.}(2010){Verth}, {Terradas}, \& {Goossens}}]{Verth2010}
{Verth}, G., {Terradas}, J., \& {Goossens}, M. 2010, \apjl, 718, L102

\bibitem[{{Woolsey} \& {Cranmer}(2015)}]{Woolsey2015}
{Woolsey}, L.~N. \& {Cranmer}, S.~R. 2015, \apj, 811, 136

\bibitem[{{Zaqarashvili} {et~al.}(2015){Zaqarashvili}, {Zhelyazkov}, \&
  {Ofman}}]{Zaqarashvili2015}
{Zaqarashvili}, T.~V., {Zhelyazkov}, I., \& {Ofman}, L. 2015, \apj, 813, 123

\end{thebibliography}

\end{document}